\documentclass[sn-mathphys,Numbered]{sn-jnl}

\usepackage{graphicx}%
\usepackage{multirow}%
\usepackage{amsmath,amssymb,amsfonts}%
\usepackage{amsthm}%
\usepackage{mathrsfs}%
\usepackage[title]{appendix}%
\usepackage{xcolor}%
\usepackage{textcomp}%
\usepackage{manyfoot}%
\usepackage{booktabs}%
\usepackage{algorithm}%
\usepackage{algorithmicx}%
\usepackage{algpseudocode}%
\usepackage{listings}%

\def\e{\epsilon}

\newcommand{\p}{{\partial}}

\usepackage{color}

\theoremstyle{thmstyleone}%
\theoremstyle{thmstyletwo}%

\theoremstyle{thmstylethree}%

\begin{document}

\title[Article Title ]{On the Effect of Liquid Crystal Orientation in the Lipid Layer on Tear Film Thinning and Breakup}



\author[1]{\fnm{M.J.} \sur{Taranchuk}}\email{mjgocken@udel.edu}

\author*[1]{\fnm{R.J.} \sur{Braun}}\email{rjbraun@udel.edu}

\affil[1]{\orgdiv{Department of Mathematical Sciences}, \orgname{University of Delaware}, \orgaddress{\city{Newark}, \state{DE},  \postcode{19716}, \country{USA}}}

\abstract{
The human tear film (TF) is thin multilayer fluid film that is critical for clear vision and ocular surface health.  Its dynamics are strongly affected by a floating lipid layer and, in health, that layer slows evaporation and helps create a more uniform tear film over the ocular surface. The tear film lipid layer (LL) may have liquid crystalline characteristics and plays important roles in the health of the tear film. Previous models have treated the lipid layer as a Newtonian fluid in extensional flow. We extend previous models to include extensional flow of a thin nematic liquid crystal atop a Newtonian aqueous layer with insoluble surfactant between them.  We derive the resulting system of  nonlinear partial differential equations for thickness of the LL and aqueous layers, surfactant transport and velocity in the LL. Evaporation is taken into account, and is affected by the LL thickness, internal arrangement of its rod-like molecules, and external conditions. Despite the complexity, this system still represents a significant reduction of the full system. We solve the system numerically via collocation with finite difference discretization in space together with implicit time stepping. We analyze solutions for different internal LL structures and show significant effect of the orientation. Orienting the molecules close to the normal direction to the TF surface results in slower evaporation, and other orientations have an effect on flow, showing that this type of model has promise for predicting TF dynamics.}





\maketitle
\section{Introduction}
\label{sec1}

As a result of each eye blink, the ocular surface is coated with thin multi-layer liquid film that typically stabilizes rapidly after each blink \cite{braun2015dynamics}. In health, the tear film helps protect the eye from foreign particles and aids clear vision \cite{willcox2017tfos}. 
The normal tear film structure may fail to form initially, or may, sometime after a blink, develop tear breakup, where the tear film fails to coat
the ocular surface \cite{EwenTBUreview17,yokoi2017classification}. Tear breakup (TBU) and associated hyperosmolarity (excessive saltiness of the local tears) is thought to play an important role in the development of dry eye disease (DED), which affects millions of people \cite{GilbardFarris78,BaudouinEtal13,craig2017tfos}.  In this work, we study a model for the dynamics for TBU in a small area of the tear film.  We study one possible model for how the lipid layer internal structure may be incorporated and affect the hyperosmolarity in TBU.

\begin{figure}[b]
\centering
\includegraphics[width=.6\textwidth]{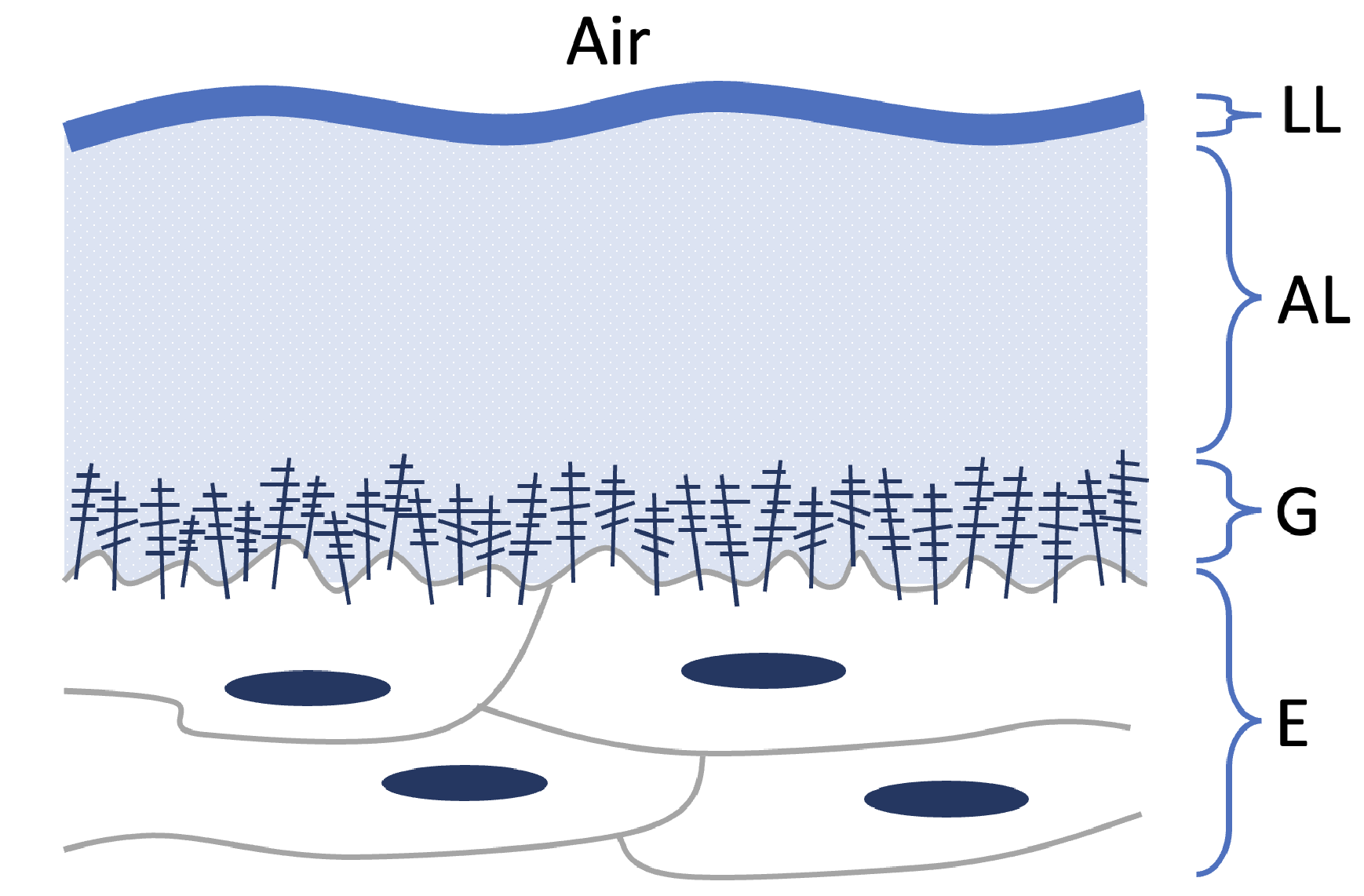}
\caption{A sketch of the tear film on the ocular surface. Here LL denotes the lipid layer, AL the aqueous layer, G the glycocalyx, and E is the outermost layer of the corneal epithelium. } \label{fig:tf_ocul_surf} 
\end{figure}

We begin with a brief primer on the tear film's structure.  
 A sketch of a cross section of a small part of the tear film is shown in Fig.~\ref{fig:tf_ocul_surf}.
Proceeding inward from air, the outermost layer is the tear film lipid layer (LL);  it averages on the 
order of tens of nanometers in thickness \cite{BronTiffRev04,KSOculSurfRev11,braun2015dynamics}.  The LL is insoluble in water and thus floats upon the next layer inward, the aqueous layer (AL).   The aqueous layer averages a few microns in thickness \cite{King-SmithFink04,jwang2003TFthickOCT}, and coats a layer of transmembrane mucins and other molecules at the ocular surface called the glycocalyx \cite{Gipson04,Gipson10,BronEtal15,fini2020membranemucins}.  Finally, the outer surface of the corneal epithelium is the beginning of the ocular surface itself \cite{hogan1971histology}. 

Simultaneous imaging of the LL and the aqueous layer \cite{King-SmithIOVS13a} shows a strong correlation between LL dynamics and TBU. The LL is typically thought to be a barrier to evaporation, thus providing an important function to preserve the tear film between blinks \cite{MishimaMaurice61,KSHinNic10}.
However, the lipid layer composition \cite{butovich2014biophysical} and structure \cite{Leiske11,Leiske12,rosenfeld2013structural} are complex and not yet fully understood. Meibum, an oily secretion from meibomian glands in the eyelids \cite{KnopKnop2011}, is the primary component of the lipid layer; it is not uncommonly used as a model for the lipid layer. X-ray scattering methods applied to \emph{in vitro} meibum films have suggested that there are ordered particles in the meibum films with layered structures \cite{rosenfeld2013structural}; these particles may have liquid crystal structure. Hot-stage imaging of meibum droplets have shown birefringence \cite{butovich2014biophysical}, another sign of order within the meibum. And in the meibomian glands \cite{KnopKnop2011} in the rat eyelid, freeze fracture with electron microscopy shows a layered structure of the lipids inside the cells that are the source of the meibum \cite{sirigu92freeze}. We interpret this evidence to suggest that the tear film lipid layer could be an extended liquid crystalline layer (possibly with defects) \cite{King-SmithOS13,paananen2019waxesters}.  It is not known whether the entire lipid layer has these qualities, or whether isolated chunks or lamellae of structured material float in the layer; however, there is general agreement that the LL has non-Newtonian properties \cite{PanNagBT99,Leiske12,rosenfeld2013structural,butovich2014biophysical,georgiev2019lipidsat}.  These areas of structure in the lipid layer are thought to provide the barrier against evaporation of the aqueous layer  \cite{rosenfeld2013structural, butovich2014biophysical,paananen2019waxesters}. 

Imaging of ocular surface shows cooling during the interblink \cite{EfYoBr89,CraigEtal00,PurWol05} and typically a correlation between TBU and tear breakup \cite{wingli2015TBUinfrared,DurschFLandThermal2017}.
Cooling of liquid crystals facilitates orientation of the molecules in the same direction \cite{YangWu15,Leiske11}. The ocular surface cools about 1 to 2 $^\circ$C during the interblink period when the eye is open \cite{EfYoBr89,CraigEtal00,PurWol05}, and is heated during the blink; thus there is thermal cycling \cite{deng2013model,deng2014heat} of the ocular surface as well as the mechanical cycling from blinks. The cooling of the lipid layer may also encourage the formation of liquid crystal structure \emph{in vivo} \cite{Leiske11,rosenfeld2013structural}.

Measurements of AL thinning rates show dependence on LL thickness \cite{king2010application}. Increased evaporation is observed at sufficiently thin LL thickness.  Typical rates for a healthy LL were around 1 to 2 $\mu$m/min, but rates increased significantly for LL thickness below 20 nm.  Measurement of thinning rates for 20 subjects and four measurements each showed a distribution of thinning rates up to about 25 $\mu$m/min \cite{king2010application}.  Thickness is not the only LL property that affects AL thinning rate \cite{fenner2015moretostable,king-smith2015moretostable}.

Turning to mathematical approaches, various mathematical models for tear film dynamics have been developed in the last 40 years or so.  
We briefly mention a sample of them proceeding from largest scale to smallest.  There have been several papers solving for flow over the open eye-shaped domain between blinks \cite{MakiBraun10a,MakiBraun10b,LiBraun14,LiBraun16,LiBraun18} and including blinking \cite{makiModelTear2019} to gain insight into overall flows and imaging of them.  Flow over whole open eye including blinking has been addressed by compartment models of the tear film \cite{GaffEtal10,CerretaniEtal14}.
Flows on one-dimensional domains spanning the open eye (vertically crossing its center) were early efforts to study tear film dynamics \cite{wong1996deposition,ShaTiwKhaTif98,MilPolRad02a,BraunFitt03,BraunUsha12}. A moving end was incorporated into modeling of tear film dynamics in one-dimensional domains in order to study a variety of effects:  deposition of the TF \cite{wong1996deposition,JonesEtal05,heryudono2007single}; insoluble surfactants representing polar lipids \cite{jones2006effect,AydemirBreward2010,maki2020influence}; heat transfer \cite{deng2013model,deng2014heat}; curvature of the ocular surface \cite{Allouche17,mehdaoui2021numerical}; the nonpolar lipid layer \cite{bruna2014influence,zubkov2012coupling}; shear thinning and eye drop design \cite{JossicLefevre09,mehdaoui2021eyedrop}.  Local models of TBU have used 1D models on a smaller region near where the tear film fails \cite{ZhangMatar03,ZhangMatar04,peng2014evaporation,BraunDrisTBU17,ZhongEtal18,ZhongJMO18,ZhongBMB19,choudhuryMembraneMucin2021,deyContinuousMucinProfiles2019,deyContinuousMucinCorrection2020}.  
Including fluorescein dye transport and fluorescence has enabled fitting of models to \textit{in vivo} fluorescence data within TBU to estimate parameters that are not possible to directly measure at this time \cite{BraunDrisTBU17,ZhongBMB19,lukeParameterEstimation2020,lukeParameterEstimationMixedMechanism2021a}.  Recently, ordinary differential equation models with no space dependence have been successfully fit to fluorescence data in small TBU spots and streaks \cite{lukeParameterEstimationMixedMechanism2021b}.  Those models have been coupled to a neural-network based data extraction system to greatly expand the amount of TBU instances that may be studied \cite{driscoll2023fittingMAIO}.

Molecular dynamics (MD) simulations of the lipid layer have also given insight into the structure and function of the lipid layer.  Coarse grained models were initially employed \cite{wizert2014organization,cwiklik2016molecular} and those models suggest that the LL may not be so as well organized as layers of nonpolar lipids atop a polar lipid monolayer.  Later, all-atom models have suggested that the specific polar lipids \cite{bland2019oahfa,viitaja2021tfll-and-oahfa} and wax esters \cite{paananen2019waxesters} may form ordered states to slow evaporation of water from the aqueous layer.  MD simulations are likely to continue to give important insights into LL structure and function, but they currently can only be solved at small space and time scales.  Those insights need to be converted to larger scale models to address other situations such as TBU.

Returning to mechanics- and differential equations-based theories, modeling of extensional flow was developed over the last century, and relevant studies for our work were summarized in \citet{taranchuk2023extensional}. Of particular interest was the work of \citet{cummings2014extensional}, who studied extensional flow of thin sheets of nematic liquid crystals in a weakly elastic limit.  They used the Ericksen-Leslie equations and multiple scale perturbation methods to reduce the equations to a small partial differential equation (PDE) system. 
In \citet{taranchuk2023extensional}, we extended their model \cite{cummings2014extensional} by rescaling the Ericksen-Leslie equations in a new limit for the case of moderate elastic effects. The scaling in \citet{taranchuk2023extensional} is based on those developed for nematic thin films on a substrate \cite{lin2013note,Lam2015}.  Both the weakly elastic and moderately elastic limits were studied for several different boundary conditions imposed at the sheet ends, and the boundary conditions were found to strongly affect the shape of the evolving sheet under stretching \cite{taranchuk2023extensional}. 

We aim to combine the aspects of a few previous models in this work.  A dynamic extensional Newtonian lipid layer was combined with a moving end and a shear-dominated Newtonian aqueous layer previously by 
\citet{bruna2014influence} and \citet{zubkov2012coupling}.  A model for TBU dynamics that used a more sophisticated approach to evaporation but a fixed LL thickness distribution was studied by \citet{peng2014evaporation}.  TBU dynamics with two Newtonian layers was studied by \citet{stapf2017duplex}. 
Here, we modify that last work \cite{stapf2017duplex} by putting a weakly-elastic nematic liquid crystal material in extension \cite{cummings2014extensional,taranchuk2023extensional} to represent the LL.  This new model will incorporate nematic molecule orientation into the evaporation resistance.  We believe that this model is the first to combine the flow of liquid crystal and Newtonian layers in this way, and also the first to incorporate not just thickness but structure of the lipid layer into the evaporation resistance.

The paper is organized as follows. In Section \ref{sec:Model} we describe the problem formulation, and in Sections \ref{sec:governing} through \ref{sec:nondim}, we present the full fluid dynamics model and scale it into nondimensional form.  In Section \ref{sec:derivation}, we derive a multiple-scale, or lubrication theory, approximation to the flow \cite{ODB97,CrasMat09review} that we will solve computationally for a sample of conditions. Section \ref{sec:Solving} provides details of the numerical method used to solve the model. In Section \ref{sec:Results} we present our results. These include profiles of the film thicknesses, velocities and pressures that lead to different outcomes depending on the internal orientation of the liquid crystal molecules. Finally, in Section \ref{sec:Discussion} we discuss the results and outline our conclusions. 

\section{Model Formulation}
\label{sec:Model}

\begin{figure}[htbp]
\centering
\includegraphics[width=.9\textwidth]{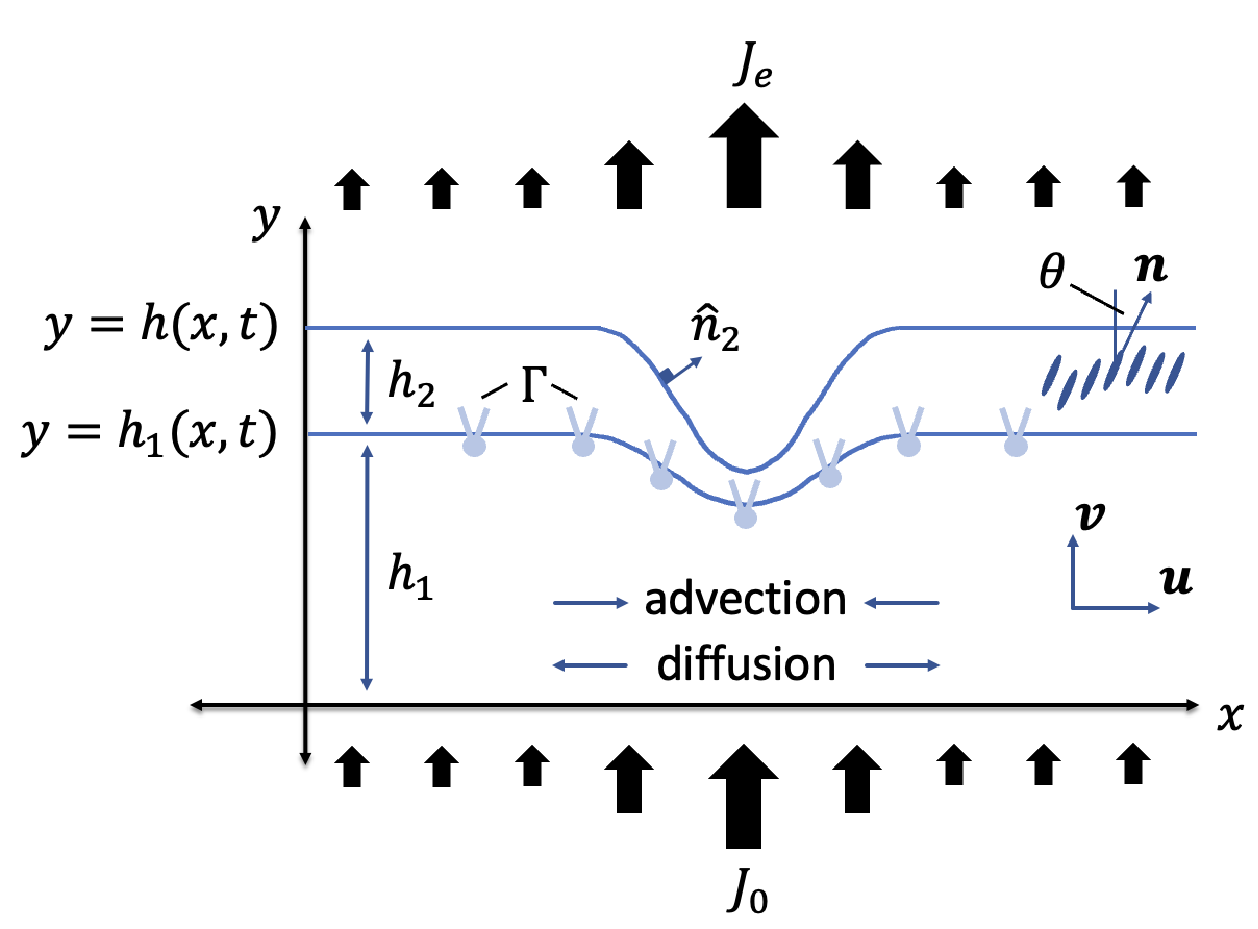}
\caption{Schematic of the two layer tear film model.}
\label{fig:schematic}
\end{figure}

We model tear film dynamics using two fluid layers. The bottom layer represents the aqueous layer of the tear film and is modeled as a Newtonian fluid, with thickness $h_1$. The top layer represents the lipid layer, and is modeled as nematic liquid crystal with weak elasticity, with thickness $h_2$. A schematic diagram of the model is shown in Figure \ref{fig:schematic}. This model accounts for loss of water through evaporation ($J_e$) and influx of water from the cornea through osmosis ($J_o$). Osmolarity, $c$, describes the concentration of salts and other solutes in the aqueous layer that induce this flow across the corneal surface. These solutes are transported in the water layer via advection and diffusion. Surfactant, largely comprising polar lipids, lies on the aqueous-lipid interface and changes the surface tension. We track the surfactant surface concentration, $\Gamma$, which is transported via advection and diffusion on that surface. Non-polar lipids are found in the LL and are assumed to be organized as a nematic liquid crystal. The director ${\bf n}$ is the preferred angle  of liquid crystal molecules through the film, as depicted in the upper right of Figure \ref{fig:schematic}.  

As we are concerned with tear film thinning and breakup, we define the TBU condition as the point at which the AL reaches a minimum thickness of $0.5\mu$m. This is roughly the height of the glycocalyx \cite{Gipson04, Gipson10}; when the aqueous layer thins to this point, the LL may rest on the top of the glycocalyx resulting in an area of dewetting \cite{EwenTBUreview17}.

\subsection{Governing equations}
\label{sec:governing}
We consider the case of two-dimensional flow in the $(x',y')$-plane. We use subscripts to indicate the fluid layer under consideration; $i=1$ corresponds to the AL, and $i=2$ corresponds to the LL. The fluid velocity is given by ${\bf u'_i} = \left(u'_i(x',y',t'), v'_i(x',y',t')\right),\;i=1,2$. Primes denote dimensional quantities. The director field, which describes the preferred orientation of the molecules relative to the $y$-axis, is ${\bf n} = \left(\sin \theta(x',y',t'), \,\cos \theta(x',y',t')\right)$. The osmotic flux of water into the domain is given by ${\bf J'_0} = (0, J'_0)$.

Let $s'_j,\;j=0,1,2$ be the position on the boundaries, with corresponding unit normal vector $\hat{n}'_j$ and unit tangent vector $\hat{t}'_j$. For example at the lipid-air interface, $y'=h'$, we have
\begin{align*}
    s'_2=(x',h'),\;\quad
    \hat{n}'_2 = \frac{1}{\sqrt{1+h^{'2}_{x'}}}(-h'_{x'},1),\;\quad
    \hat{t}'_2 = \frac{1}{\sqrt{1+h^{'2}_{x'}}}(1,h'_{x'}).
\end{align*}
At the aqueous-lipid interface, $h'_1$ replaces $h'$. 
At the corneal-aqueous interface, $y'=0$, we assume the cornea is flat.
\begin{align*}
    s'_0=(x',0),\;\quad
    \hat{n}'_0 = (0,1),\;\quad
    \hat{t}'_0 = (1,0).
\end{align*} 

\subsubsection{Aqueous layer}
Inside the AL, $0<y'<h'_1$, we have conservation of mass and momentum using the Navier-Stokes equations. The advection-diffusion equation governs osmolarity transport within the AL. The equations, respectively, are:
\begin{align}
    \nabla' \cdot {\bf u'}_1=0,\\
    \rho_1 ({\bf u'}_{1t}+{\bf u'}_1\cdot \nabla' {\bf u'}_1)= -\nabla' p'_1 + \mu_1 \Delta' {\bf u'}_1,\\
    c'_{t'} + {\bf u'}_1\cdot\nabla' c' - D_1 \Delta' c'=0,
\end{align}
where the divergence of ${\bf u'_1}$ is $\nabla'\cdot{\bf u'_1}=u'_{1x'}+v'_{1y'}$, the gradient is $\nabla'{\bf u'_1}=\begin{bmatrix} u'_{1x'} &v'_{1x'}\\u'_{1y'}& v'_{1y'}\end{bmatrix}$, and the Laplacian of the first component of ${\bf u'_1}$ is $\Delta' u'_1=u'_{1x'x'}+u'_{1y'y'}$. 
\subsubsection{Lipid layer}
Inside the LL, $h'_1<y'<h'$, we have conservation of energy, mass, and momentum using the Ericksen-Leslie equations for liquid crystals. As liquid crystals are influenced by electric fields, the complete Ericksen-Leslie equations include this effect \cite{cummings2014extensional}. In a biological setting such as the eye this is not relevant, and so we do not include those effects in the equations below. 
\begin{align}
    \frac{\partial }{\partial x'_i}\left( \frac{\partial W'}{\partial \theta_{x'_i}}\right)-\frac{\partial W'}{\partial \theta}+\tilde{g}'_i\frac{\partial n_i}{\partial \theta}&=0,  \\
    -\frac{\partial \pi'}{\partial x'_i}+\tilde{g}'_k\frac{\partial n_k}{\partial x'_i}+\frac{\partial \tilde{t}'_{ik}}{\partial x'_k}&=0,  \\
    \frac{\partial u'_i}{\partial x'_i}&=0.
\end{align}
These equations are defined fully below: 
\begin{align}
    \frac{\partial }{\partial x'}\left( \frac{\partial W'}{\partial \theta_{x'}}\right)+\frac{\partial }{\partial y'}\left( \frac{\partial W'}{\partial \theta_{y'}}\right)-\frac{\partial W'}{\partial \theta}+\tilde{g}'_x\frac{\partial n_x}{\partial \theta}+\tilde{g}'_y\frac{\partial n_y}{\partial \theta}&=0,\\
    -\frac{\partial \pi'}{\partial x'}+\tilde{g}'_x\frac{\partial n_x}{\partial x'}+\tilde{g}'_y\frac{\partial n_y}{\partial x'}+\frac{\partial \tilde{t}'_{xy}}{\partial y'}+\frac{\partial \tilde{t}'_{xx}}{\partial x'}&=0,\\
    -\frac{\partial \pi'}{\partial y'}+\tilde{g}'_x\frac{\partial n_x}{\partial y'}+\tilde{g}'_y\frac{\partial n_y}{\partial y'}+\frac{\partial \tilde{t}'_{yx}}{\partial x'}+\frac{\partial \tilde{t}'_{yy}}{\partial y'}&=0,\\
    \frac{\partial u_2'}{\partial x'}+\frac{\partial v_2'}{\partial y'}&=0,
\end{align}
where, working in two dimensions, 
\begin{align}
    \tilde{g}'_i=&-\gamma_r N'_i-\gamma_t e'_{ik}n_k, \hspace{25pt} e'_{ij}=\frac{1}{2}\left(\frac{\partial u'_i}{\partial x'_j}+\frac{\partial u'_j}{\partial x'_i}\right), \\
    N'_i=&\;\dot{n}'_i-\omega'_{ik}n_k, \hspace{55pt} \omega'_{ij}=\frac{1}{2}\left(\frac{\partial u'_i}{\partial x'_j}-\frac{\partial u'_j}{\partial x'_i}\right),\hspace{25pt}\pi'=\;p_2'+W',\\
    W'=& \;\frac{1}{2}\bigg[K_1(\nabla' \cdot {\bf n})^2+K_3(({\bf n}\cdot \nabla') {\bf n})\cdot (({\bf n}\cdot \nabla') {\bf n})\bigg],\\
    \tilde{t}'_{ij}=&\;\alpha_1'n_kn_pe'_{kp}n_in_j+\alpha'_2N'_in_j+\alpha'_3 N'_jn_i+\alpha'_4e'_{ij}+\alpha'_5e'_{ik}n_kn_j+\alpha'_6e'_{jk}n_kn_i .
\end{align}
Summation over repeated indices is understood, with $i,j,k=1,2$, and $\dot{n}_i$ denotes the convective derivative of the $i$th component of ${\bf n}$. We use summation notation here as it is standard when working with the Ericksen-Leslie equations. We further assume that the elastic constants are equal; that is $K=K_1=K_3$. These liquid crystal quantities are defined further in Table \ref{tab:lc_param}. 

\begin{table}[ht]
\caption{Liquid crystal variables and parameters \cite{stewart2019static}. }
\label{tab:lc_param}%
\begin{tabular}{ll}
\toprule
Quantity  & Description\\
\midrule
${\bf n}= (\sin\theta,\cos\theta)$& director field\\
$\theta(x',y',t')$&angle the director angle makes with the $y$-axis\\
$p'_2$      & pressure in LL \\
$W'$      & bulk energy density \\
$\pi'=p'_2+W'$& modified pressure\\
$\tilde{g}'_i$& viscous dissipation \\
$e'_{ij}$ & rate of strain tensor\\
$N'_i$ & co-rotational time flux of the director \bf{n}\\
$\omega'_{ij}$ & vorticity tensor \\
$\tau^{V'}_{2ij}$& viscous stress tensor\\
$\tau^{E'}_{2ij}$& elastic stress tensor\\
$\tau_2'$ & stress tensor\\
$\alpha'_i,\,i=1,...,6$ & Leslie viscosities (Newtonian: $\mu'=\alpha'_4/2$, all other $\alpha_i=0$)\\
$\gamma'_r=\alpha'_3-\alpha'_2$ & rotational/twist viscosity \\
$\gamma'_t=\alpha'_6-\alpha'_5$ & torsion coefficient \\
$K_1,\;K_3$      & elastic constants for splay and bend respectively    \\
\botrule
\end{tabular}
\end{table}

\subsubsection{Boundary conditions}
\paragraph{Aqueous-corneal interface}
At $y'=0$, we have velocity continuity and a no flux boundary condition for osmolarity transport. These are, respectively:
\begin{align}
    {\bf u'_1} = {\bf J_0},\\
     c'\,({\bf u'}_1 - {\bf s'}_{0t'})-D_1 \nabla' c'=0.
\end{align}
Note that as we assume the cornea is flat, ${\bf s'}_{0t'}={\bf 0}$.
\paragraph{Aqueous-lipid interface}
At $y'=h'_1$, we have velocity continuity, aqueous and lipid mass conservation, and the stress balance. In addition, we have a no flux boundary condition for osmolarity transport, and the anchoring boundary condition for the LL. The equations, respectively, are:
\begin{align}
    ({\bf u'}_1 - {\bf u'}_2)\cdot \hat{t}'_1 = 0,\\
    \rho_1 ({\bf u'}_1 - {\bf s'}_{1t'}) \cdot \hat{n}'_1 = J'_e,\\
    \rho_2 ({\bf s'}_{1t'}-{\bf u'}_2) \cdot \hat{n}'_1 = 0,\\
    ({\bf \tau'}_1 - {\bf \tau'}_2)\cdot \hat{n}'_1 = -\gamma'_{s1}\hat{n}'_1 \nabla' \cdot \hat{n}'_1 + \nabla'_{s_1} \gamma'_{s1},\\
    c'\,({\bf u'}_1 - {\bf s'}_{1t'})-D_1 \nabla' c'=0,\label{eq:ALLLnoflux}\\
    \theta=\theta_B.
\end{align}
To resolve the stress balance into tangential and normal stress components, we take the dot product of the stress balance and $\hat{t}'_1$ or $\hat{n}'_1$ respectively. Here $\nabla'_{s_1}=({\bf I}-\hat{n}'_1\hat{n}'_1)\cdot \nabla'$ is the surface gradient, where ${\bf I}$ is the identity matrix \cite{stone1990simple}. 
\paragraph{Lipid-air interface}
At $y'=h'$, we have lipid mass conservation, stress balance, and anchoring respectively:
\begin{align}
    \rho_2 ({\bf u'}_2 - {\bf s'}_{2t'})\cdot \hat{n}'_2=0,\\
    {\bf \tau'}_2 \cdot \hat{n}'_1=-\gamma_2 \hat{n}'_2 \nabla' \cdot \hat{n}'_2,\\
    \theta=\theta_B.
\end{align}
The stress balance can be resolved into components in the same way as above, using the unit tangent and normal vectors at this interface, $\hat{t}'_2$ and $\hat{n}'_2$. 

\subsubsection{Surfactant transport}
At the aqueous-lipid interface $y'=h'_1$, we have surfactant transport and the linear equation of state for the surface tension, respectively:
\begin{align}
    \Gamma'_{t'}+\nabla'_{s_1}\cdot(\Gamma' u'_1) = D_s \nabla^{'2}_{s_1}\Gamma',\\
    \gamma'_{s1}=\gamma_1 - RT_0\Gamma'.\label{eq:eqofstate}
\end{align}

\subsubsection{Osmosis}

The effect of osmosis is determined by the difference in concentration on either side of the aqueous-corneal interface $y'=0$,
\begin{align}
    {\bf J'}_0 \cdot \hat{n}'_0 &= P_c\,(c'-c_0).
\end{align}

\subsubsection{Evaporation}
To account for the effect of evaporation at the aqueous-lipid interface $y'=h'_1$, we modify a boundary layer model derived thoroughly in \citet{stapf2017duplex}
 \begin{align}
    J'_e&=\frac{E_0}{1+\frac{k_m}{Dk}h'_2}.
\end{align}
Here, $Dk$ represents resistance to evaporation provided by the LL based on its thickness and permeability. Its value comes from a nonlinear least squares fit to clinical data \cite{ king2010application} performed by \citet{stapf2017duplex}, which is very similar to that of \citet{cerretani2013water}. We further modify $Dk$ to include a dependence on the director angle of liquid crystals in the LL. We suggest the orientation of the molecules affects evaporation through 
\begin{align}
    J'_e&=\frac{E_0}{1+\frac{k_m}{Dk}(0.1+0.9\sin \theta_B)h'_2}.
\end{align}
The maximum evaporation rate occurs when $\theta_B=0$, which is ten times that of the rate for $\theta_B=\pi/2$.  The minimum rate matches that used in \cite{stapf2017duplex}.

\subsection{Scalings}
\label{sec:scale}
We non-dimensionalize the model using the  following scalings:

\begin{align}
    x'&=Lx, \qquad\qquad  h_1'=\e L h_1,\qquad\qquad h_2'=\e \delta L h_2,\nonumber\\
y'&=\e Ly,\hspace{36pt}      u_1'=Uu_1, \hspace{44pt}      u_2'= Uu_2,  \nonumber\\
h'&=\e L h, \hspace{36pt}     v_1'=\e U v_1,\hspace{42pt} v_2'=\e U v_2,  \nonumber\\
t'&=\frac{L}{U}t,\hspace{38pt} p_1'= \frac{\mu_1 U}{\e^2L}p_1,\hspace{33pt} p_2'= \frac{\mu_2 U}{L}p_2,  \\
 \Gamma'& = \Gamma_0\Gamma,\hspace{38pt} c'=Cc,\hspace{46pt}  W'=\frac{K}{\delta^2 L^2}W, \nonumber\\
 J_o'& = \e U J_o,\hspace{32pt} \gamma'_t=\mu_2\gamma_t,\hspace{43pt}     \alpha_i'=\mu_2\alpha_i\nonumber,\\
 J_e'&=\e\rho_1 UJ_e,\hspace{22pt}\gamma'_r=\mu_2\gamma_r.\nonumber
\end{align}

Here $h'=h'_1+h'_2\implies h=h_1+\delta h_2$. The parameters used in the scalings are given in Table \ref{tab:scalings}, and the non-dimensional parameters are listed in Table \ref{tab:nondim}. 

We have chosen parameter values and initial conditions corresponding to those in \citet{stapf2017duplex} in order to compare results from their base case scenario. The model from \citet{stapf2017duplex} represents the lipid layer as a Newtonian fluid with evaporation given as in Equation (\ref{eq:evap}). We model the lipid layer as liquid crystal, and hypothesize a way in which the orientation of the liquid crystal molecules may affect evaporation. However, these liquid crystal parameters have not been measured in the human eye, and so exactly what values the parameters should take is unknown. In general, there are only a handful of liquid crystals whose properties have been measured. In the realm of liquid crystals, we return to a Newtonian fluid by setting the Leslie viscosities $\alpha'_i=0,\,i=1,2,3,5,6$; $\alpha'_4=2\mu_2$. The dynamic viscosity of the lipid layer, $\mu_2$, has previously been taken to be 0.1 Pa s \cite{bruna2014influence, stapf2017duplex}. Thus, to match this viscosity in the Newtonian limit, we choose to set our Leslie viscosities to be three times the properties of 5CB, a relatively well studied liquid crystal \cite{stewart2019static}; see Table \ref{tab:physparam} for the specific values as well as the values of other physical parameters.

\begin{table}[htbp]
\caption{Parameters used in model scalings} \label{table:scalings}%
\begin{tabular}{@{}lllll@{}}
\toprule
	  Parameter
		&  Value 
		&  Units
		&  Description
		&  Reference  \\ 
\midrule
	 $L=H_1(\gamma_1+\gamma_2)/(\mu_1 E_0)^{1/4}$
		& $3.1779 \times 10^{-4}$
		& m
		& Length scale
		&   \cite{stapf2017duplex} \\
	 $H_1$
		& $3.5 \times 10^{-6}$
		& m
		& AL thickness 
            & \cite{king2004thickness}\\
	 $H_2$
		& $5 \times 10^{-8}$
		& m
		& LL thickness
		&  \cite{KSHinNic10}  \\
	$V = E_0$
            & $5.093\times 10^{-7}$ 
            & m/s
            & Thinning rate
            &\cite{stapf2017duplex}\\
	$U=V/\e $
		& $4.624\times 10^{-5}$
		& m/s
		& Horizontal velocity
		& \cite{stapf2017duplex}   \\
	$t=L/U$
		& $6.8726$
		& s
		& Time scale
		&  Calculated  \\
	 $C$
		& $300$
		& mOsM
		& Reference osmolarity
		& \cite{peng2014evaporation}   \\
	 $\Gamma_0$
		& $4 \times 10^{-7}$
		& mol/$\text{m}^2$
		& Surfactant concentration
		& \cite{AydemirBreward2010,bruna2014influence}  \\
\botrule
\end{tabular}
\label{tab:scalings}
\end{table}

\begin{table}[htbp]
\caption{Nondimensional Parameters} 
\label{table:base_parameters}%
\begin{tabular}{@{}llll@{}}
\toprule
	Parameter
		&  Formula
		&  Value 
		&  Description  \\  
\midrule
	$\epsilon$
		&  $H_1 / L$
		&  $0.0110$
		&  Aspect ratio of AL  \\
	$\delta$
		&  $H_2 / H_1$
		&  $0.0143$
		&  Lipid to aqueous thickness ratio \\
	$\Upsilon$
		&  $\epsilon^2 \mu_2 / \mu_1$
		&  $0.0091$
		&  Reduced viscosity ratio  \\
	$\mathcal{C}_1$
		&  $\epsilon^3 (\gamma_1+\gamma_2) / (\mu_1 U)$
		&  $1.0001$
		&  Reduced aqueous capillary number  \\
	$\mathcal{C}_2$
		&  $\epsilon \gamma_2 / (\mu_2 U)$
		&  $43.8371$
		&  Reduced lipid capillary number  \\
	$\mathcal{M}$
		&  $\epsilon R T_0 \Gamma_0 / (\mu_1 U)$
		&  $187.7687$
		&  Reduced Marangoni number  \\
	$\mathcal{E}$
		&  $ E_0 / (\epsilon\rho_1  U)$
		&  $1.0001$
		&  Evaporation parameter  \\
	$\mathcal{R}_0$
		&  $k_m H_2 / (Dk)$
		&  $17.68$
		&  Evaporative resistance of LL  \\
	$\mathcal{P}$
		&  $C P_c / (\epsilon U)$
		&  $0.1355$
		&  Osmosis parameter  \\
    $\mbox{Re}_1$       
            &   $\rho_1 UL / \mu_1$      
            &   0.0113    
            &   Reynold's number for AL          \\
	$\textrm{Pe}_1$
		& $UL/D_1$
		&  $9.1842$
		&  P\'{e}clet number for osmolarity diffusion  \\
	$\textrm{Pe}_s$
		&  $UL/D_s$
		&  $0.4898$
		&  P\'{e}clet number for surfactant diffusion  \\
\botrule
\end{tabular}
\label{tab:nondim}
\end{table}

\begin{table}[ht]
\caption{Physical Parameters}
\label{table:constants}%
\begin{tabular}{@{}lllll@{}}
\toprule
	Constant
		&  Value 
		&  Units
		&  Description
		&  Reference  \\  
\midrule
	$\rho_1$
		&  1000
		&  kg/$\text{m}^3$
		&  Density of liquid water
		&  \cite{CRC77} \\
	$\rho_2$
		&  900
		&  kg/$\text{m}^3$
		&  Density of lipid
		&  \cite{bruna2014influence} \\
	$\mu_1$
		&  $1.3 \times 10^{-3}$
		&  Pa s
		&  Dynamic viscosity of water
		&  \cite{bruna2014influence,tiffany1991viscosity}  \\
	$\mu_2=\alpha_4'/2$
		&  $0.0978$
		&  Pa s
		&  Dynamic viscosity of lipid
		&  \cite{stewart2019static} \\
	$\gamma_1$
		&  $0.027$
		&  N/m
		&  Surface tension of water
		&	 \cite{tiffany1987lipid}	\\
	$\gamma_2$
		&  $0.018$
		&  N/m
		&  Surface tension of lipid
		&	 \cite{nagyova1999components}	\\
	$R$
		&  $8.3145$
		&  J/(K mol)
		&  Ideal gas constant
		&  \cite{CRC77}	\\
	$D_1$
		&  $1.6 \times 10^{-9}$
		&  $\text{m}^2$/s
		&  Osmolarity diffusion constant
		&  \cite{riquelme2007interferometric}  \\
	$D_s$
		&  $3 \times 10^{-8}$
		&  $\text{m}^2$/s
		&  Surfactant diffusion constant
		&  \cite{bruna2014influence}  \\
	$k_m$
		&  $0.0182$
		&  $ \text{m} / \text{s}$
		&  Mass transfer coefficient
		&  \cite{stapf2017duplex}  \\
	$Dk$
		&  $ 5.136  \times 10^{-11}$
		&  $ \text{m}^2 / \text{s}$
		&  Lipid permeability
		&  \cite{stapf2017duplex}  \\
	$P_c$
		&  $2.3 \times 10^{-10}$
		&  $\text{kg}/(\text{mOsM} \ \text{m}^2 \ \text{s})$
		&  Corneal osmosis coefficient
		&  \cite{peng2014evaporation}  \\
	$E_0$
		&  $5.0928 \times 10^{-7}$
		&  $ \text{m} / \text{s}$
		&  Evaporative thinning rate
		&  \cite{stapf2017duplex}  \\
        $T_0$
        & 308.15
        & K
        & Eye surface temperature
        &\cite{peng2014evaporation} \\
        $\alpha_1'$
            & $-0.0180$
            & Pa s
            & Leslie viscosity 
            & \cite{stewart2019static}\\
        $\alpha_2'$
            & $-0.2436$
            & Pa s
            & Leslie viscosity
            & \cite{stewart2019static}\\
        $\alpha_3'$
            & $-0.0108$
            & Pa s
            & Leslie viscosity
            & \cite{stewart2019static}\\
        $\alpha_4'$
            & $0.1956$
            & Pa s
            & Leslie viscosity
            & \cite{stewart2019static}\\
        $\alpha_5'$
            & $0.1920$
            & Pa s
            & Leslie viscosity
            & \cite{stewart2019static}\\
        $\alpha_6'$
            & $-0.0624$
            & Pa s
            & Leslie viscosity
            & \cite{stewart2019static}\\
        $\gamma_r' = \alpha_3'-\alpha_2'$ 
            & 0.2328
            &Pa s
            & Rotational viscosity
            & \cite{stewart2019static}\\
        $\gamma_t'= \alpha_6'-\alpha_5'$ 
            & 0.1296
            &Pa s
            & Torsion coefficient 
            & \cite{stewart2019static}\\
        $\theta_B$
            &$0$ - $\pi/2$
            & radians
            & Director angle
            & \cite{stewart2019static}\\
\botrule
\end{tabular}
\label{tab:physparam}
\end{table}

\subsection{Nondimensional equations}
\label{sec:nondim}
\subsubsection{Aqueous layer}
Inside the AL, $0<y<h_1$, the conservation of mass and momentum equations become
\begin{align}
    u_{1x}+v_{1y}=0,\\
    \e^2 (u_{1t}+u_1u_{1x}+v_1u_{1y}) = \frac{1}{\mbox{Re}_1}(-p_{1x} + \e^2 u_{1xx} + u_{1yy}),\\
    \e^4 (v_{1t}+u_1v_{1x}+v_1v_{1y}) = \frac{1}{\mbox{Re}_1}(-p_{1y} + \e^4 v_{1xx} + \e^2 v_{1yy}),
\end{align}
and the osmolarity transport equation becomes
\begin{align}
    \e^2 \left(c_t+c_xu_1+c_yv_1-\frac{1}{\mbox{Pe}_1}c_{xx}\right)-\frac{1}{\mbox{Pe}_1}c_{yy}&=0.
\end{align}

\subsubsection{Lipid layer}
In the LL, $h_1<y<h$, the Ericksen-Leslie equations become
\begin{align}
    \e^2 \hat{N}\frac{\p}{\p x}\left(\frac{\partial W}{\p \theta_x}\right)+\hat{N}\frac{\p}{\p y}\left(\frac{\partial W}{\p \theta_y}\right)+ \tilde{g}_x\frac{\p n_x}{\p \theta}+ \tilde{g}_y\frac{\p n_y}{\p \theta}&=0,\\
    -\e^2\frac{\partial p_2}{\partial x}-\e\hat{N}\frac{\partial W}{\p x}+\e \tilde{g}_x\frac{\p n_x}{\p x}+\e \tilde{g}_y\frac{\p n_y}{\p x}+\e\frac{\p \tau_{xx}}{\p x}+\frac{\p \tau_{xy}}{\p y}&=0,\\
    -\e\frac{\partial p_2}{\partial y}-\hat{N}\frac{\partial W}{\p y}+ \tilde{g}_x\frac{\p n_x}{\p y}+ \tilde{g}_y\frac{\p n_y}{\p y}+\e\frac{\p \tau_{yx}}{\p x}+\frac{\p \tau_{yy}}{\p y}&=0,\\
    \frac{\p u_2}{\p x}+\frac{\p v_2}{\p y}&=0.
\end{align}

\subsubsection{Boundary conditions}
\paragraph{Aqueous-corneal interface}
At $y=0$, velocity continuity and the normal component of the no flux boundary condition become, respectively:
\begin{align}
    u_1=0,\\
    v_1=J_o,\\
    \e^2 cv_1-\frac{1}{\mbox{Pe}_1}c_y=0.
\end{align}
\paragraph{Aqueous-lipid interface}
At $y=h_1$, velocity continuity gives 
\begin{align}
   \left(1+\e^2h_{1x}^2\right)^{-1/2} \left(u_1-u_2+\e^2[v_1-v_2]h_{1x}\right)=0,
\end{align}
aqueous and lipid mass conservation respectively give
\begin{align}
    \left(1+\e^2h_{1x}^2\right)^{-1/2}(h_{1t}+u_1h_{1x}-v_1)+J_e &=0,\\
   \left(1+\e^2h_{1x}^2\right)^{-1/2}\left( h_{1t}+u_2h_{1x}-v_2\right)&=0.
\end{align}
The stress balance becomes
\begin{align}
    \left(1+\e^2h_{1x}^2\right)^{-1/2}\bigg[ (\e {\bf \tau}_1 -\Upsilon {\bf \tau}_2) \left(-\e h_{1x},1\right)\bigg]=\nonumber\\
    \left(1+\e^2h_{1x}^2\right)^{-2} (\mathcal{C}_1-\e^2\mathcal{M}\Gamma) h_{1xx} \left(-\e^2h_{1x},\,\e \right)-\e^2\mathcal{M}\nabla_{s_1}\Gamma,
\end{align}
where we have used the nondimensional version of Equation (\ref{eq:eqofstate}), and 
\begin{align}
    \nabla_{s_1}=(1+\e^2h_{1x}^2)^{-1}\begin{bmatrix}
        \partial_x+h_{1x}\partial_y\\
        \e h_{1x}\partial_x+\e h_{1x}^2\partial_y
    \end{bmatrix}.
\end{align}
Finally, the normal component of the no flux boundary condition for osmolarity transport and the anchoring boundary condition respectively become
\begin{align}
   \left(1+\e^2h_{1x}^2\right)^{-1/2}\left[-\e^2cu_1h_{1x}+\e^2 c(v_1-h_{1t})-\frac{1}{\mbox{Pe}_1}\left(c_y-\e^2 c_xh_{1x}\right)\right]=0,\\
    \theta=\theta_B.
\end{align}
\paragraph{Lipid-air interface}
At the lipid-air interface $y=h$, we have lipid mass conservation, stress balance, and anchoring
\begin{align}
   \left(1+\e^2h_{1x}^2\right)^{-1/2}\left(h_t+u_2h_x-v_2\right)=0,\\
    \left(1+\e^2h_{1x}^2\right)^{-1/2}\bigg[{\bf \tau}_2 (-\e h_x,1)\bigg]= \left(1+\e^2h_{x}^2\right)^{-2} \left(-\e^2 \mathcal{C}_2h_xh_{xx},\e \mathcal{C}_2h_{xx}\right),\\
    \theta=\theta_B.
\end{align}
\paragraph{Surfactant transport}
Surfactant transport at the aqueous-lipid interface $y=h_1$ becomes
\begin{align}
     \Gamma_t+ \nabla_{s_1}\cdot(\Gamma {\bf u}_1)= \frac{1}{\mbox{Pe}_s}\nabla^2_{s_1}\Gamma.
\end{align}
\paragraph{Osmosis}
The equation for osmosis at $y=0$ becomes
\begin{align}
    J_o=P\left(c-\frac{c_0}{C}\right).
\end{align}
\paragraph{Evaporation}
The equation for the evaporation at $y=h_1$ becomes
\begin{align}
    J_e=\frac{\mathcal{E}}{1+\mathcal{R}(\theta_B)h_2},
\end{align}
where
\begin{align}
   \mathcal{R}(\theta_B)=\mathcal{R}_0(0.1+0.9\sin \theta_B).\label{eq:thetadep_evap}
\end{align}
Then, $\theta_B=\pi/2$, describing a vertical orientation of the molecules, gives the default evaporative resistance of $\mathcal{R}=\mathcal{R}_0=17.68$, while an angle of $\theta_B<\pi/2$ reduces the evaporative resistance, with $\theta_B=0$, describing a horizontal orientation of the molecules, reducing resistance by a factor of 10.

\subsection{Model derivation}
\label{sec:derivation}
Now we asymptotically expand all the dependent variables in powers of the small parameter $\epsilon \ll 1$. For example,
\[\Gamma(x,t) \sim \Gamma_0(x,t)+\e\, \Gamma_1(x,t)+\e^2 \Gamma_2(x,t)+\cdots\]

Note that $h_1,\,h_2,\,h$, and $\Gamma$ are functions of $x$ and $t$ only. Initially, all other variable expansions are functions of $x,\,y$, and $t$ unless they are determined to be independent of one or more variables. We substitute these expansions into the nondimensionalized equations and collect like powers of $\e$. 

\subsubsection{Leading order}
At leading order, the equations are greatly simplified. Working outward from the cornea toward the air, the equations are as follows.\\
At the corneal-aqueous interface $y=0$,
\begin{align}
    u_{10}&=0, \label{eq:velcontu1}\\
    v_{10}&=J_o, \label{eq:velcontv1}\\
    \frac{1}{\mbox{Pe}_1}c_{0y}&=0,\label{eq:ALnofluxy}
\end{align}
with osmosis given by
\begin{align}
    J_o&=P(c_{0}-1).
\end{align}
Inside the aqueous layer, $0<y<h_1$,
\begin{align}
   u_{10x}+ v_{10y}=0, \label{eq:ALincompressibility}\\
    \frac{1}{\mbox{Re}_1}(p_{10x}-u_{10yy})=0,\label{eq:ALxmomentum}\\
    \frac{1}{\mbox{Re}_1}p_{10y}=0, \label{eq:ALymomentum}\\
    \frac{1}{\mbox{Pe}_1}c_{0yy}=0.\label{eq:salttransport}
\end{align}
At the aqueous-lipid interface $y=h_1$,
\begin{align}
    u_{10}-u_{20}=0, \label{eq:ALLLvelcont}\\
    h_{10t}+u_{10}h_{10x}-v_{10}=-J_e, \label{eq:ALmasscons}\\
    h_{10t}+u_{20}h_{10x}-v_{20}=0, \label{eq:ALLLmasscons}\\
    \frac{\Upsilon}{4}\bigg[-4+\alpha_2-\alpha_3-\alpha_5-\alpha_6+(\alpha_2+\alpha_3-\alpha_5+\alpha_6)\cos(2\theta_0)\nonumber\\ +\frac{\alpha_1}{2}(\cos(4\theta_0)-1)\bigg]u_{20y}=0,\label{eq:ALLLstressbalance1}\\
    -\frac{\Upsilon}{4}\bigg[\alpha_1+\alpha_2+\alpha_3+\alpha_5+\alpha_6+\alpha_1\cos(2\theta_0)\bigg]\sin(2\theta_0)u_{20y}+\hat{N}\Upsilon\theta_{0y}^2=0, \label{eq:ALLLstressbalance2}\\
    \frac{1}{\mbox{Pe}_1}c_{0y}=0, \label{eq:ALLLnofluxy}\\
    \theta_0-\theta_B=0, \label{eq:ALLLanchoring}
\end{align}
with surfactant transport given by
\begin{align}
    \Gamma_{0t}+(\Gamma_0u_{10})_x =\frac{1}{\mbox{Pe}_s}\Gamma_{0xx}. \label{eq:surfactanttransport}
\end{align}
Inside the lipid layer, $h_1<y<h$,
\begin{align}
    \frac{1}{2}\left[-\alpha_2+\alpha_3+(\alpha_5-\alpha_6)\cos(2\theta_0)\right]u_{20y}+\hat{N}\theta_{0yy}&=0,\label{eq:energy}\\
    \frac{1}{2}\left[\alpha_2+\alpha_3-\alpha_5+\alpha_6+2\alpha_1\cos(2\theta_0)\right]\sin(2\theta_0)u_{20y}\theta_{0y}\nonumber\\-\frac{1}{4}\left[-4+2(\alpha_2-\alpha_5)\cos(\theta_0)^2-2(\alpha_3+\alpha_6)\sin(\theta_0)^2-\alpha_1\sin(2\theta_0)^2\right]u_{20yy}&=0, \label{eq:LLxmomentum}\\
    \frac{1}{2}\left[-\alpha_2+\alpha_3+(\alpha_1+\alpha_2+\alpha_3+2\alpha_5)\cos(2\theta_0)+\alpha_1\cos(4\theta_0)\right]u_{20y}\theta_{0y}\nonumber\\+\frac{1}{4}\left[\alpha_1+\alpha_2+\alpha_3+\alpha_5+\alpha_6+\alpha_1\cos(2\theta_0)\right]\sin(2\theta_0)u_{20yy}-3\hat{N}\theta_{0y}\theta_{0yy}&=0, \label{eq:LLymomentum}\\
    u_{20x}+v_{20y}&=0. \label{eq:LLincompressibility}
\end{align}
At the lipid-air interface $y=h$,
\begin{align}
    h_{0t}+u_{20}h_{0x}-v_{20}&=0,\label{eq:LLmasscons}\\
   -\frac{1}{4}\bigg[-4+\alpha_2-\alpha_3-\alpha_5-\alpha_6+(\alpha_2+\alpha_3-\alpha_5+\alpha_6)\cos(2\theta_0)\nonumber\\
   +\frac{\alpha_1}{2}(\cos(4\theta_0)-1)\bigg]u_{20y}&=0, \label{eq:LLstressbalance1}\\
    \frac{1}{4}\bigg[\alpha_1+\alpha_2+\alpha_3+\alpha_5+\alpha_6+\alpha_1\cos(2\theta_0)\bigg]\sin(2\theta_0)u_{20y}-\hat{N}\theta_{0y}^2&=0, \label{eq:LLstressbalance2}\\
    \theta_0 - \theta_B &=0, \label{eq:LLanchoring}
\end{align}
with evaporation given by
\begin{align}
    J_e&=\frac{\mathcal{E}}{1+\mathcal{R}(\theta_B)h_{20}}. \label{eq:thinningrate}
\end{align}
\subsubsection{Solving the leading order equations}
Subtracting the product of $\theta_{0y}$ and Equation (\ref{eq:energy}) from Equation (\ref{eq:LLymomentum}) and simplifying results in the following derivative of products
 \begin{align} \label{eq:82}
     \bigg(\frac{1}{4}\left[\alpha_1+\alpha_2+\alpha_3+\alpha_5+\alpha_6+\alpha_1\cos(2\theta_0)\right]\sin(2\theta_0)u_{20y}-2\hat{N}\theta_{0y}^2\bigg)_y&=0.
 \end{align}
Thus, the inner functions must be independent of $y$. Evaluating Equation (\ref{eq:82}) at $y=h_1$ and applying Equation (\ref{eq:ALLLstressbalance1}), yields $\theta_{0y}=D_0(x,t)$. Thus, $\theta_{0yy}=0$, and substituting this into Equation (\ref{eq:energy}), yields $u_{20y}=0$. Applying the boundary conditions from Equations (\ref{eq:ALLLstressbalance2}) and (\ref{eq:ALLLanchoring}), gives
\begin{align}
u_{20}=u_{20}(x,t)\quad \text{and}\quad \theta_0=\theta_B.
\end{align}
That is, we have extensional flow in the lipid layer that is independent of $y$ at leading order, as in \citet{stapf2017duplex}, and the angle of the molecules is constant throughout the depth and in time at leading order, as in \citet{cummings2014extensional}.

Integrating Equation (\ref{eq:ALincompressibility}) with respect to $y$, and applying the boundary conditions from Equations (\ref{eq:velcontv1}) and (\ref{eq:ALmasscons}) conserves mass of water for the AL:
\begin{align}
    h_{10t}+(\Bar{u}_{10}h_{10})_x&=J_o-J_e.\label{eq:lubrication}
\end{align}   
We now turn to the transverse velocity in the lipid layer. Integrating Equation (\ref{eq:LLincompressibility}) with respect to $y$ and applying the boundary condition at $y=h_1$ from Equation (\ref{eq:ALLLmasscons}) yields
\begin{align}
    v_{20}&=-u_{20x}y+h_{10t}+u_{20}h_{10x}+u_{20x}h_{10}.
\end{align}
Evaluating this equation $y=h$, using the boundary condition in Equation (\ref{eq:LLmasscons}) we have
\begin{align}
    h_{20t}+\left(u_{20}h_{20}\right)_x&=0.
\end{align}
This is conservation of mass for the lipid layer. 

\subsubsection{\texorpdfstring{$O(\epsilon)$}{Lg} equations}
We proceed to the next order to solve for $u_{21}$. We take the $x$-momentum equation in the lipid layer at $O(\e)$, and simplify using Equation (\ref{eq:LLincompressibility}). This gives
\begin{align}
    u_{21yy}&=0.
\end{align}
Integrating twice with respect to $y$ and applying the $O(\e)$ tangential stress boundary condition at $y=h$ and continuity of velocity at $y=h_1$ yields
\begin{align}
    u_{21}=&-\frac{\left[\alpha_5-\alpha_6-\alpha_1\cos(2\theta_B)\right]\sin(2\theta_B)}{2+(\alpha_5-\alpha_2)\cos^2(\theta_B)+[\alpha_1+\alpha_3+\alpha_6+\alpha_1\cos(2\theta_B)]\sin^2(\theta_B)}u_{20x} (y-h_{10}) \nonumber\\
    &+ u_{11}(x,h_{10},t).
\end{align}
We now seek to determine $u_{10}$. 
The $O(\e)$ normal stress condition at $y=h_1$ is
\begin{align}
\left[-4\alpha_1\cos(\theta_B)^3\sin(\theta_B)-(\alpha_2+\alpha_3+\alpha_5+\alpha_6)\sin(2\theta_B)\right]\Upsilon u_{21y}\nonumber\\
+\left[-8-4(\alpha_5+\alpha_6)\cos(\theta_B)^2-4\alpha_1\cos(\theta_B)^4\right]\Upsilon v_{20y}\nonumber\\
-\alpha_1\Upsilon\sin(2\theta_B)^2u_{20x}-4p_{10}+4\Upsilon p_{20}-4\mathcal{C}_1h_{10xx}&=0.\label{eq:shearstresse}
\end{align}
Solving Equation (\ref{eq:shearstresse}) for $p_{10}$, at $y=h_1$,
\begin{align}
        p_{10} =&\Upsilon\bigg[\frac{1}{8}A_1-\frac{A_2}{A_3} u_{20x}+ p_{20}\bigg]_{y=h_1}-\mathcal{C}_1h_{10xx}.\label{eq:p10}
\end{align}
where
\begin{align}
A_1=&16+3\alpha_1+4(\alpha_5+\alpha_5)+4(\alpha_1+\alpha_5+\alpha_6)\cos(2\theta_B)\nonumber\\
&+\alpha_1\cos(4\theta_B)-2\alpha_1\sin^2(2\theta_B),\label{eq:A1}\\
A_2=&2(\alpha_6-\alpha_5+\alpha_1\cos(2\theta_B))(\alpha_1+\alpha_2+\alpha_3+\alpha_5+\alpha_6\nonumber\\
&+\alpha_1\cos(2\theta_B))\sin^2(2\theta_B),\\
A_3=&2+(\alpha_5-\alpha_2)\cos^2(\theta_B)+(\alpha_1+\alpha_3+\alpha_6+\alpha_1\cos(2\theta_B))\sin^2(\theta_B).
\end{align}

\subsubsection{\texorpdfstring{$O(\epsilon^2)$}{Lg} equations}
We also need the shear stress equation at the aqueous-lipid interface at $O(\e^2)$. Both Equation (\ref{eq:p10}) and the 
 shear stress condition involve functions that need to be evaluated at $y=h_1$ and functions that are independent of $y$. We write them as:
\begin{align}
    p_{10}=\Upsilon F(x,t) - \mathcal{C}_1 h_{10xx}\label{eq:p10simp},\\
    u_{10y}=\Upsilon G(x,t) -\mathcal{M}\Gamma_{0x}\label{eq:u10y},
\end{align}
where
\begin{align}
    F(x,t)=\bigg[\frac{1}{8}A_1-\frac{A_2}{A_3} u_{20x}+ p_{20}\bigg]_{y=h_1},
\end{align}
\begin{align}
    G(x,t)=\;&\frac{1}{8} \bigg[2 \bigg(B_1u_{20x}+2 \left[\alpha_1 \cos (2 \theta_B)
    +\alpha_2+\alpha_3\right]\sin (2 \theta_B) u_{21y}+B_2v_{20y}\bigg)h_{10x}\nonumber\\
    &-2  [\alpha_1 \cos (2 \theta_B)-\alpha_1-2 \alpha_5]\sin (2 \theta_B) u_{21x}-B_3u_{22y}-B_4v_{20x}\nonumber\\
    &+2 [\alpha_1 \cos (2 \theta_B)+\alpha_1+2 \alpha_6]\sin (2 \theta_B) v_{21y}\bigg]_{y=h_1}.
\end{align}
where
\begin{align}
B_1=&\,2  (\alpha_1+\alpha_5+\alpha_6)\cos (2 \theta_B)-\alpha_1 \cos (4 \theta_B)-\alpha_1-2 \alpha_5-2 \alpha_6-8,\\
B_2=&\,2  (\alpha_1+\alpha_5+\alpha_6)\cos (2 \theta_B)+\alpha_1 \cos (4 \theta_B)+\alpha_1+2 \alpha_5+2 \alpha_6+8,\\
B_3=&\,\alpha_1 \cos (4 \theta_B)-\alpha_1+2  (\alpha_2+\alpha_3-\alpha_5+\alpha_6)\cos (2 \theta_B)+2 \alpha_2\nonumber\\
&-2 \alpha_3-2 \alpha_5-2 \alpha_6-8,\\
B_4=\,&\alpha_1 \cos (4 \theta_B)-\alpha_1-2  (\alpha_2+\alpha_3+\alpha_5-\alpha_6)\cos (2 \theta_B)-2 \alpha_2\nonumber\\
&+2 \alpha_3-2 \alpha_5-2 \alpha_6-8.
\end{align}
Then at $y=h_1$,
\begin{align}
    p_{10x}= \Upsilon F'(x,t) - \mathcal{C}_1 h_{10xxx}. \label{eq:p10x}
\end{align}
Integrating Equation (\ref{eq:ALxmomentum}) with respect to $y$ twice and applying Equations (\ref{eq:p10x}), (\ref{eq:u10y}),  and (\ref{eq:velcontu1}) gives
\begin{align}
    u_{10}&=p_{10x}\bigg(\frac{y^2}{2}-h_{10}y\bigg)+\bigg(\Upsilon G(x,t)-\mathcal{M}\Gamma_{0x}\bigg)y.\label{eq:u10}
\end{align}
Now, the depth averaged axial velocity at leading order is
\begin{align}
    \Bar{u}_{10}&=\frac{1}{h_1}\int_0^{h_{10}}u_{10}\,dy\\
    &= -p_{10x}\frac{h_{10}^2}{3}+\bigg(\Upsilon G(x,t)-\mathcal{M}\Gamma_{0x}\bigg)\frac{h_{10}}{2}.
\end{align}
In order to solve for $p_{20}$, we use the $O(\e)$ equation for $y$-momentum in the lipid layer,
\begin{align}
    \frac{1}{8}\bigg[16+3\alpha_1+4\alpha_5+4\alpha_6+4(\alpha_1+\alpha_5+\alpha_6)\cos(2\theta_B)+\alpha_1\cos(4\theta_B)\bigg]v_{20yy}\nonumber\\
    +\frac{1}{4}\bigg[\alpha_1+\alpha_2+\alpha_3+\alpha_5+\alpha_6+\alpha_1 \cos(2\theta_B)\bigg]\sin(2\theta_B)u_{21yy}-p_{20y}=0.\label{eq:p20y}
\end{align}
We will also need the normal stress equation at the lipid-air interface $y=h$ at $O(\e)$,
\begin{align}
    -p_{20}+\frac{1}{4}\alpha_1\sin(2\theta_B)^2u_{20x}-\mathcal{C}_2h_{0xx}\nonumber\\
    +\bigg[\alpha_1\cos(\theta_B)^3\sin(\theta_B)+\frac{1}{4}(\alpha_2+\alpha_3+\alpha_5+\alpha_6)\sin(2\theta_B)\bigg]u_{21y}\nonumber\\
    +\bigg[2+(\alpha_5+\alpha_6)\cos(\theta_B)^2+\alpha_1\cos(\theta_B)^4\bigg]v_{20y}&=0.\label{eq:p2bc}
\end{align}
Integrating Equation (\ref{eq:p20y}) with respect to $y$ and applying Equation (\ref{eq:p2bc}) gives
\begin{align}
    p_{20}=&\frac{1}{4}\bigg[\alpha_1+\alpha_2+\alpha_3+\alpha_5+\alpha_6+\alpha_1 \cos(2\theta_B)\bigg]\sin(2\theta_B)u_{21y}\nonumber\\
    &+\frac{1}{8}H_1v_{20y}+\alpha_1\cos(\theta_B)^2\sin(\theta_B)^2u_{20x}-\mathcal{C}_2h_{0xx},
\end{align}
where
\begin{align}
H_1=16+3\alpha_1+4\alpha_5+4\alpha_6+4(\alpha_1+\alpha_5+\alpha_6)\cos(2\theta_B)+\alpha_1\cos(4\theta_B).
\end{align}
Substituting the solution for $p_{20}$ simplifies the complicated expression for $F(x,t)$ to
\begin{align}
    F(x,t)=-\mathcal{C}_2h_{0xx}.
\end{align}
Substituting $F(x,t)$ into the equation for $p_{10}$ shows us that the pressure in the aqueous layer is independent of $y$.

The next step is to find an expression for $u_{20}$. This involves working with equations from $O(\e^2)$ which are too lengthy to write down in their entirety here. Integrating the $x$-momentum equation in the lipid layer at $O(\e^2)$ with respect to $y$, and solving for $u_{22y}$ yields
\begin{align}
    u_{22y}=M(\theta_B,x,t)y+K(x,t),\label{eq:u22y}
\end{align}
where $M$ is lengthy, and $K(x,t)$ is the constant of integration \cite{taranchuk2023extensional}. Applying the $O(\e)$ shear stress condition at the lipid-air interface determines $K(x,t)$. We substitute Equation (\ref{eq:u22y}) into Equation (\ref{eq:u10}) for $u_{10}$. We evaluate this equation at the aqueous-lipid interface, and apply continuity of velocity at $O(1)$, which gives an equation in terms of $u_{20}$.  We find
\begin{align}
   -\frac{A(\theta_B)}{B(\theta_B)} \delta\Upsilon (u_{20x} h_{20})_x  -\frac{u_{20}}{h_{10}}=&\frac{h_{10}}{2} \bigg(-\mathcal{C}_1 h_{10xxx}- \mathcal{C}_2 \Upsilon h_{0xxx}\bigg)\nonumber\\
   &- \mathcal{C}_2\delta\Upsilon h_{20} h_{0xxx} +\mathcal{M} \Gamma_{0x},
\end{align}
where
\begin{align}
A(\theta_B)=&\,2 \cos (2 \theta_B) (\alpha_1+\alpha_5+\alpha_6+4) (\alpha_2+\alpha_3-\alpha_5+\alpha_6)\nonumber\\
&+\cos (4 \theta_B) (\alpha_1 [\alpha_2-\alpha_3]+[\alpha_2+\alpha_3] [\alpha_5-\alpha_6])+\alpha_1 \alpha_2-\alpha_1 \alpha_3-2 \alpha_1 \alpha_5\nonumber\\
&-2 \alpha_1 \alpha_6-8 \alpha_1+\alpha_2 \alpha_5+3 \alpha_2 \alpha_6+8 \alpha_2-3 \alpha_3 \alpha_5-\alpha_3 \alpha_6-8 \alpha_3-2 \alpha_5^2\nonumber\\
&-4 \alpha_5 \alpha_6-16 \alpha_5-2 \alpha_6^2-16 \alpha_6-32,\\
B(\theta_B)=&-\alpha_1 \cos (4 \theta_B)+\alpha_1-2 \cos (2 \theta_B) (\alpha_2+\alpha_3-\alpha_5+\alpha_6)\nonumber\\
&-2 \alpha_2+2 \alpha_3+2 \alpha_5+2 \alpha_6+8.
\end{align}
Note that if $\alpha_i=0,\;i\neq 4$, we recover the Newtonian case \cite{bruna2014influence,stapf2017duplex},
\begin{align}
4\delta\Upsilon(h_{20}u_{20x})_x-\frac{u_{20}}{h_{10}}=&\frac{h_{10}}{2}  \bigg(-\mathcal{C}_1 h_{10xxx}
-\mathcal{C}_2 \Upsilon h_{0xxx}\bigg)\nonumber\\&-\mathcal{C}_2 \delta \Upsilon h_{20} h_{0xxx}+\mathcal{M} \Gamma_{0x}.
\end{align}
At this point, we can simplify $G(x,t)$. After substitution, 
\begin{align}
G(x,t)&= \mathcal{C}_2\delta h_{20} h_{0xxx}- \frac{A(\theta_B)}{B(\theta_B)}\delta ( h_{20}u_{20x} )_x.
\end{align}

\subsubsection{Finding the equations for surfactant and osmolarity transport}
The final steps are to solve the equations for surfactant surface concentration and osmolarity. 
To find the leading order equation for the surfactant concentration, we apply continuity of velocity, Equation (\ref{eq:ALLLvelcont}), and rearrange to obtain
\begin{align}
   \Gamma_{0t}+ (\Gamma_0 u_{20})_x=\frac{1}{\mbox{Pe}_s} \Gamma_{0xx}.
\end{align}
The leading order equations for osmolarity, Equations (\ref{eq:ALnofluxy}), (\ref{eq:salttransport}), and (\ref{eq:ALLLnofluxy}) reveal that $c_0=c_0(x,t)$. To find an explicit equation for $c_0(x,t)$ requires the $O(\e)$ and $O(\e^2)$ equations. At $O(\e)$,
\begin{align}
    c_{1y}&=0,\quad\text{at }\,y=0,\\
    c_{1yy}&=0\quad\text{for }\,0<y<h_1,\\
    c_{1y}&=0,\quad\text{at }\,y=h_1.
\end{align}
At $O(\e^2)$,
\begin{align}
    c_0 J_o-\frac{1}{\mbox{Pe}_1}c_{2y}&=0\quad\text{at }\,y=0,\label{eq:cyat0}\\
    c_{0t}-\frac{1}{\mbox{Pe}_1}c_{2yy}+u_{10}c_{0x}-\frac{1}{\mbox{Pe}_1}c_{0xx}&=0\quad\text{for }\,0<y<h_1,\label{eq:salttrans}\\
    c_{0x}h_{10x}+\mbox{Pe}_1c_0J_e-c_{2y}&=0\quad\text{at }\,y=h_1.\label{eq:cyath1}
\end{align}
Here we have used Equations (\ref{eq:velcontv1}) and (\ref{eq:ALmasscons}) from leading order to substitute for $v_{10}$ at the boundaries, as well as the tangential component of Equation (\ref{eq:ALLLnoflux}). 
Solving Equation (\ref{eq:salttrans}) for $c_{0yy}$ and integrating with respect to $y$ gives
\begin{align}
   c_{2y}&= \bigg[\left(\mbox{Pe}_1c_{0t}-c_{0xx}\right)y\bigg]_0^{h_1} +\mbox{Pe}_1c_{0x}\int_0^{h_1}u_{10}\,dy.
\end{align}
We now evaluate this expression at both boundaries, applying Equations (\ref{eq:cyat0}) and (\ref{eq:cyath1}) and using Equation (\ref{eq:lubrication}). We find
\begin{align}
    (c_0h_{10})_t &=\frac{1}{\mbox{Pe}_1}(c_{0x}h_{10})_x-\left(c_0\Bar{u}_{10}h_{10}\right)_x.
\end{align}
This completes the derivation of the PDEs governing AL thickness, LL velocity and thickness, surfactant concentration, and osmolarity.

\subsubsection{The model system}
\label{sec:modelsys}
After simplification, the closed system of equations that we wish to solve is
\begin{align}
    h_{10t}+(\Bar{u}_{10}h_{10})_x=J_o-J_e,\\
    h_{20t}+\left(u_{20}h_{20}\right)_x=0,\\
    p_{10}=-\Upsilon\mathcal{C}_2h_{0xx} - \mathcal{C}_1 h_{10xx},\label{eq:pressure1}\\
    -\frac{A(\theta_B)}{B(\theta_B)} \delta\Upsilon (u_{20x} h_{20})_x -\frac{u_{20}}{h_{10}}=\frac{h_{10}}{2} p_{10x}
- \mathcal{C}_2\delta\Upsilon h_{20} h_{0xxx} +\mathcal{M} \Gamma_{0x},\label{eq:axforcebal}\\
\Gamma_{0t}+ (\Gamma_0 u_{20})_x=\frac{1}{\mbox{Pe}_s}\Gamma_{0xx},\\
m_{0t} =\frac{1}{\mbox{Pe}_1}\left(m_{0x}-\frac{m_0h_{10x}}{h_{10}}\right)_x-\left(m_0\Bar{u}_{10}\right)_x,
\end{align}
where
\begin{align}
    h_0=h_{10}+\delta h_{20},\\
    J_0=P\left(\frac{m_0}{h_{10}}-1\right),\\
    J_e= \frac{\mathcal{E}}{1+\mathcal{R}(\theta_B)h_{20}},\label{eq:evap}\\
    \Bar{u}_{10}=-p_{10x}\frac{h_{10}^2}{12}+\frac{u_{20}}{2}.\label{eq:ubar}
\end{align}
We have reduced the order of the system by adding $p_1$ as a dependent variable. We also use the substitution $m_0=h_{10}c_0$, where $m_0$ is the mass.

\subsubsection{Initial and boundary conditions}
\label{sec:ic-and-bc}
The boundary conditions are homogeneous Neumann conditions on $h_1$, $h_2$, $p_1$, $\Gamma$, and $c$ to enforce no flux of these variables.  We specific $u_2=0$ for the lipid layer velocity. \\
We choose initial conditions in such away that a local defect in the lipid layer, a thin spot, drives all the dynamics of the tear film toward TBU. We choose a Gaussian perturbation of the lipid layer 
\begin{align}
    h_{2}(x,0)=1-0.9e^{-\frac{x^2}{2\sigma^2}},
\end{align}
where $2\sigma^2=1/9$.
 We take $h_1=\Gamma=c=1$ initially. We compute $p_1(x,0)$ from Equation~(\ref{eq:pressure1}), and solve the discrete version of the axial force balance Equation~(\ref{eq:axforcebal}) using the other initial values to obtain $u_2(x,y,0)$. 

\subsection{Numerical solution method}
\label{sec:Solving}
We solve the system of equations in Section \ref{sec:modelsys}, subject to the initial and boundary conditions of Section \ref{sec:ic-and-bc}, numerically using the method of lines. Spatial derivatives are approximated with second-order finite differences on a uniform grid of $512$ points. The result is a system of differential algebraic equations that we solve forward in time in \textsc{Matlab} (MathWorks, Natick, MA, USA) using \texttt{ode15s}. We use event detection to stop the simulation if the AL thickness reaches a minimum value of $0.5\mu$m, which we consider to represent TBU.  We found that the solutions were converged using this number of grid points, and that the mass of water and osmolarity were conserved to the tolerances of the computation.  

We integrate the system until either: (i) the time at which TBU occurs (TBU time, or TBUT) and that becomes the final time; or (ii) a final time of 60 s is reached.  Whether TBU is reached or not depends on the parameters of the particular computation.

\section{Results}
\label{sec:Results}
\subsection{Base case}
\begin{figure}[htbp]%
\includegraphics[width=0.49\textwidth]{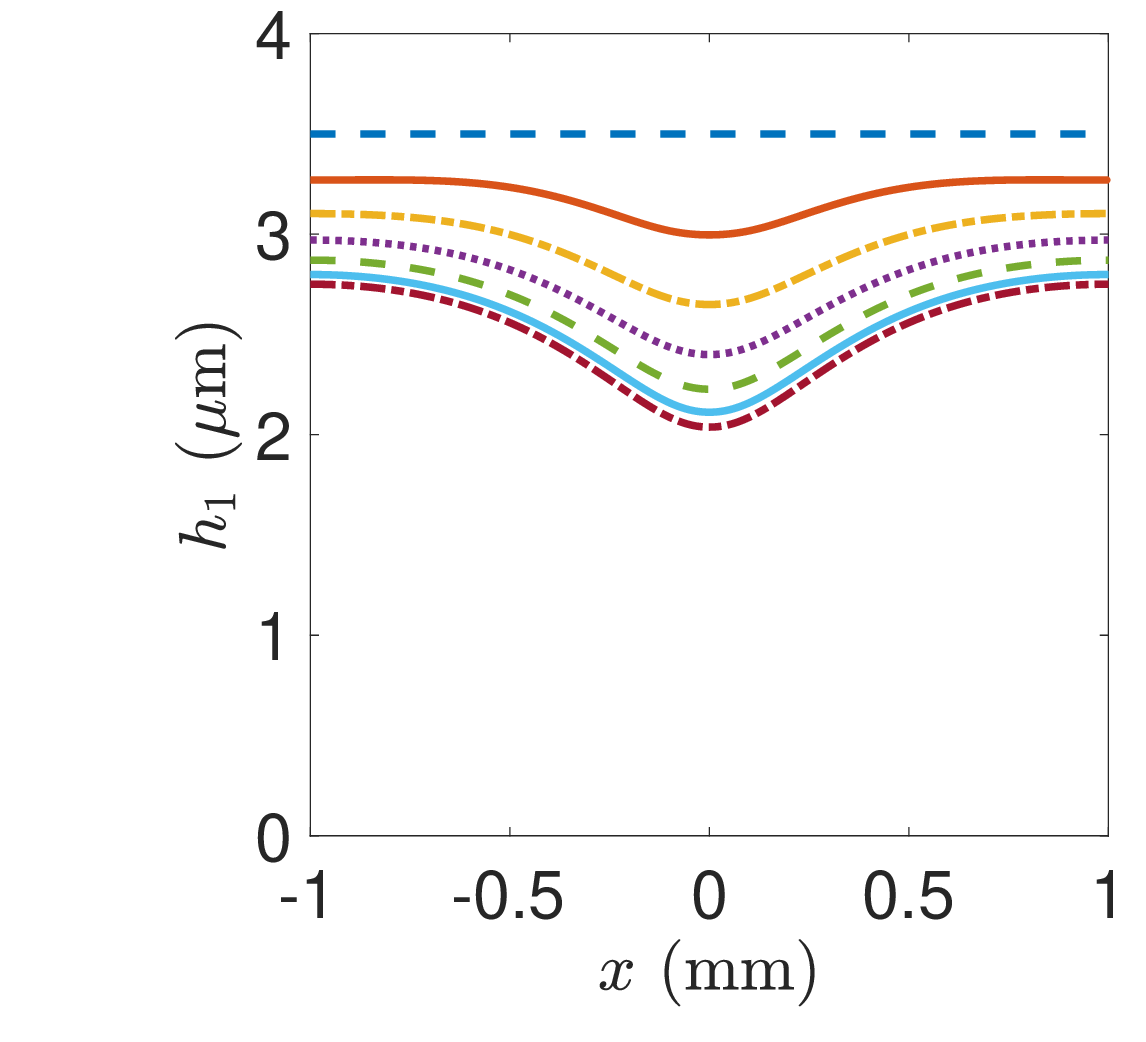}
\includegraphics[width=0.49\textwidth]{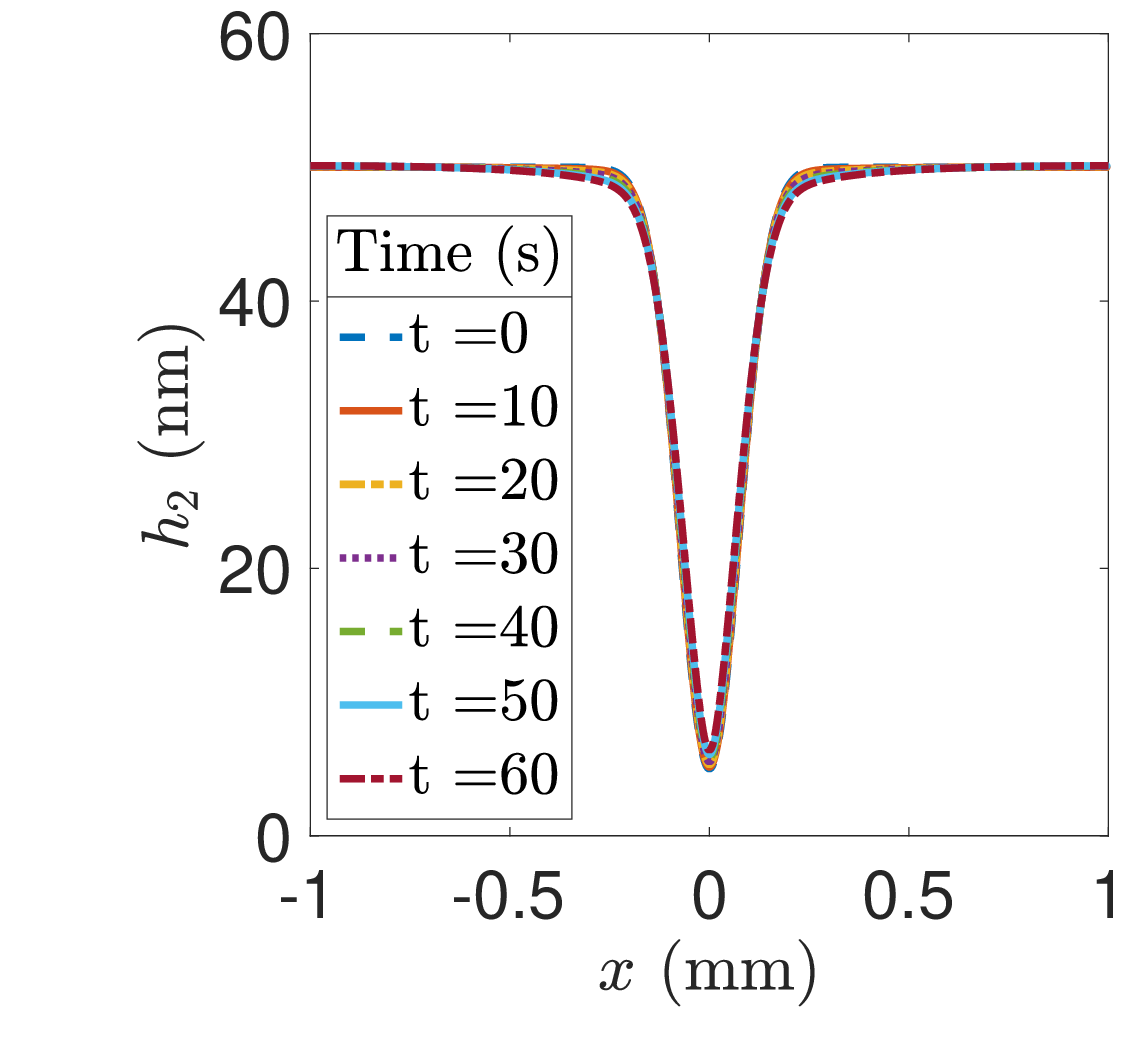}
\includegraphics[width=0.49\textwidth]{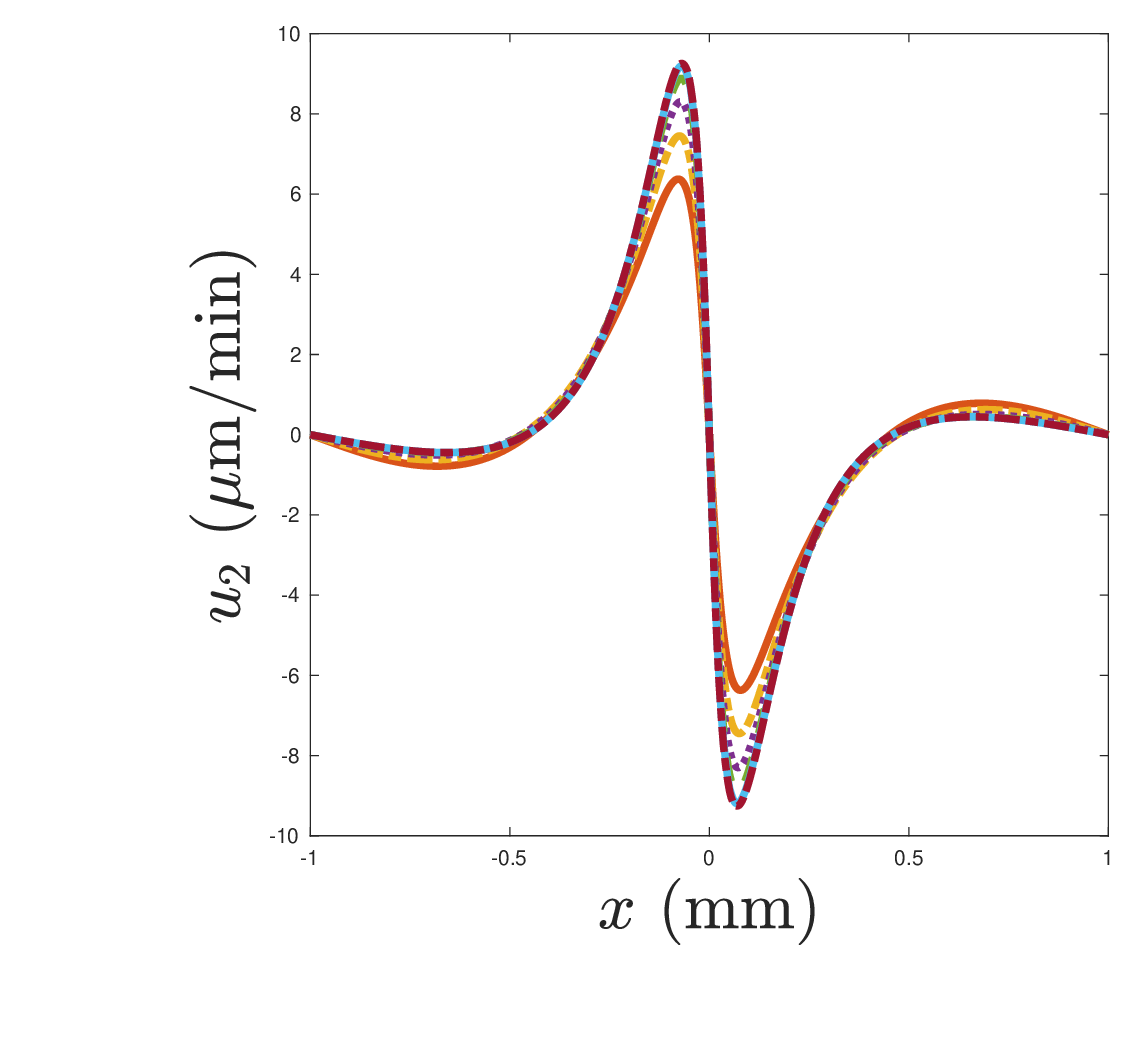}
\includegraphics[width=0.49\textwidth]{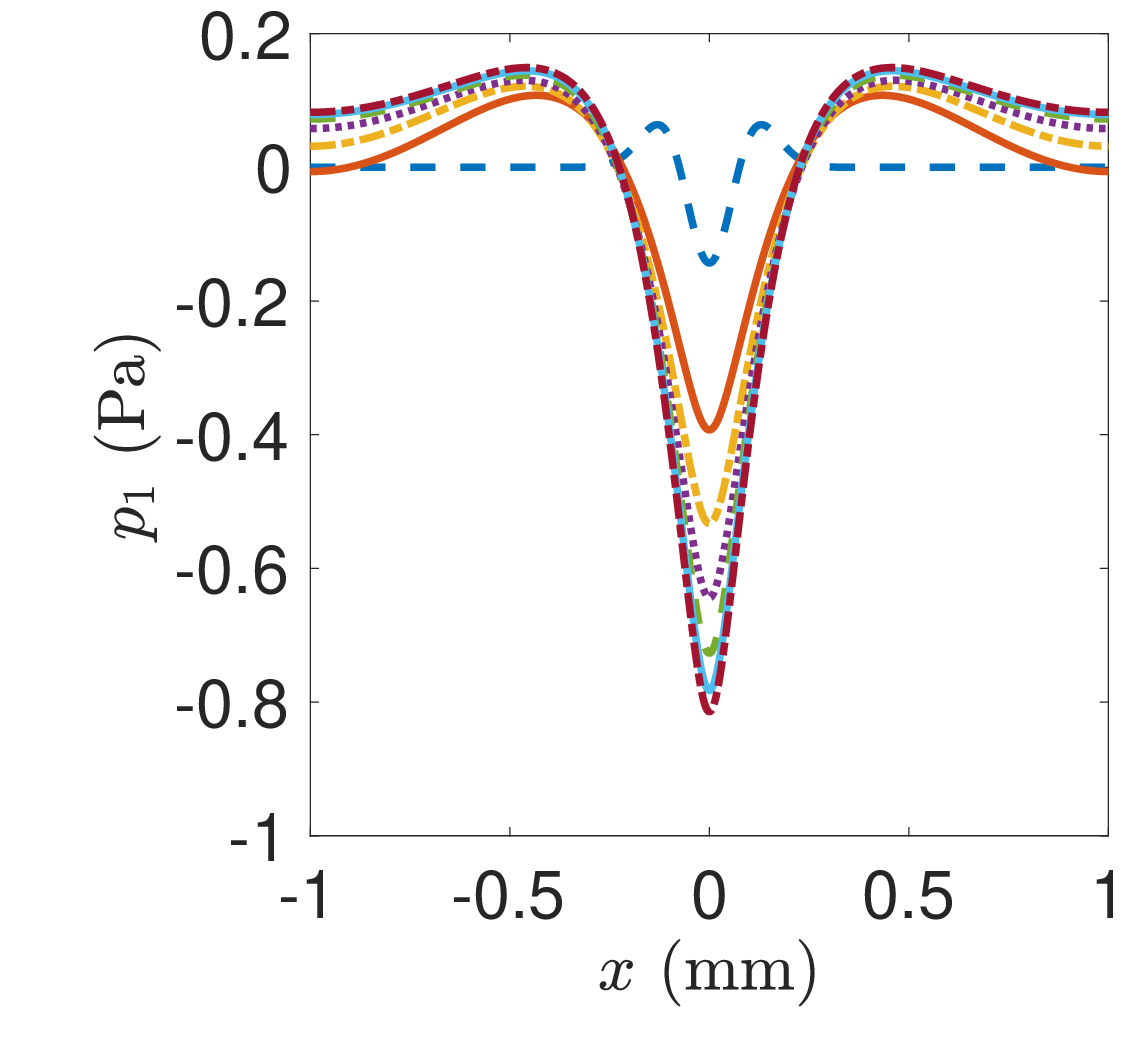}
\includegraphics[width=0.49\textwidth]{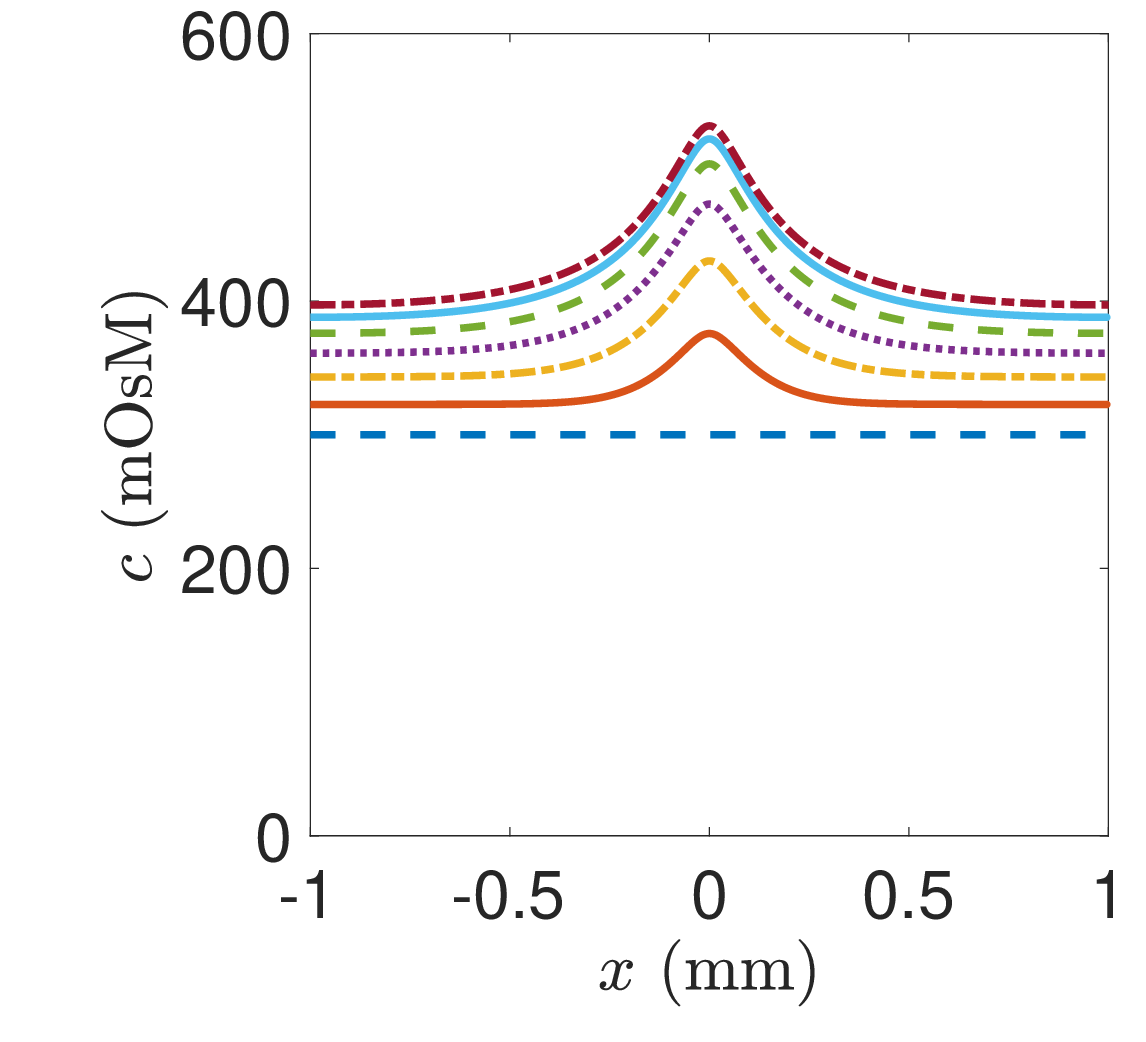}
\includegraphics[width=0.49\textwidth]{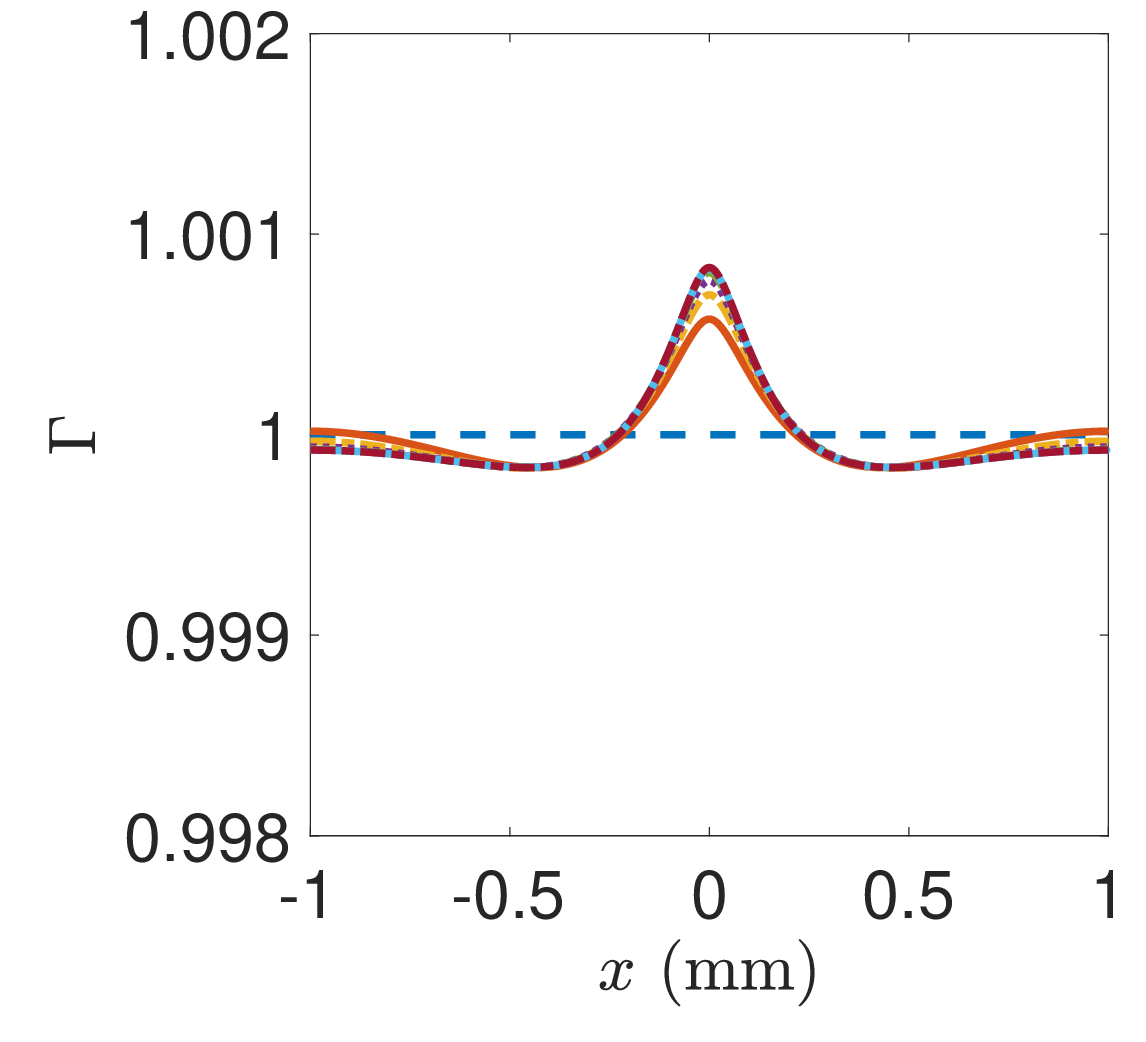}
\caption{\label{fig:basecase}Solutions for spatially-uniform initial conditions except for a Gaussian disturbance in the lipid layer thickness. The director angle of $\theta_B=\pi/2$. This figure illustrates typical dynamics of the dependent variables in space and time. The increased evaporation from the thin lipid disturbance drives all the dynamics.}
\end{figure}
For the base case, we use the $\theta$-dependent evaporation model given in Equation (\ref{eq:thetadep_evap}) and parameter values given in Table \ref{tab:physparam}. Note that by using the default value of $\theta_B=\pi/2$, we have full LL evaporation resistance; that is $\mathcal{R}=\mathcal{R}_0$. This scenario reproduces the base case of \citet{stapf2017duplex} (their Figure 3). 

Representative solutions are shown in Figure \ref{fig:basecase}. They are typical of the underlying dynamics for tear film thinning due to a localized thin spot in the initial LL profile. The LL changes very little over time from its initial shape. We see the effect of elevated local evaporation from the AL, which thins the AL at the same location.  The thinning rate slows as time increases because the local drop in pressure in the center of the domain draws flow of the AL inward due to capillarity (``healing flow" in \cite{peng2014evaporation}). The inward flow counteracts the thinning effect of evaporation to some degree but does not stop thinning \cite{peng2014evaporation,stapf2017duplex,BraunDrisTBU17}.  

The surfactant surface concentration $\Gamma$, shown relative to its initial value, builds up slightly in the middle.  Increased $\Gamma$ decreases the surface tension, which induces a shear stress pulling fluid out of the middle of the domain.  This latter (Marangoni) effect opposes the capillarity-driven flow from the pressure difference \cite{peng2014evaporation,ZhongBMB19,lukeParameterEstimationMixedMechanism2021a};  however, in all the simulations for this paper, the pressure-driven flow is the larger contribution to the AL flow. 

The osmolarity increases with increasing time, primarily centrally. The increase results from evaporation of water, leaving the osmolarity behind.  The increase is exacerbated by the inward capillary flow  that advects osmolarity to the center, but is mitigated by diffusion of osmolarity outward from the locally high concentration. 

\subsection{Effect of liquid crystal orientation}

\begin{figure}[htbp]%
\centering
\includegraphics[width=0.49\textwidth]{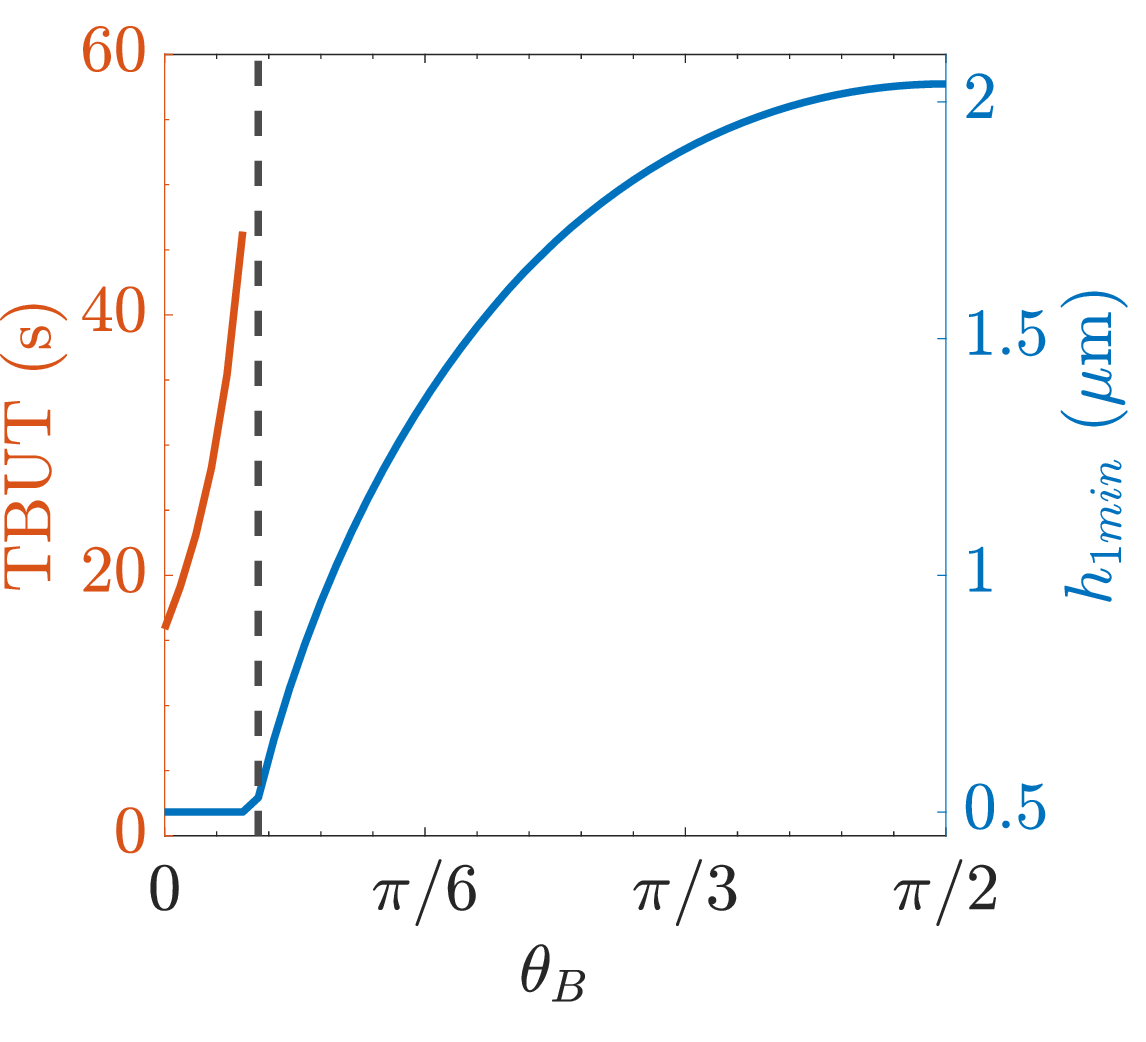}
\includegraphics[width=0.49\textwidth]{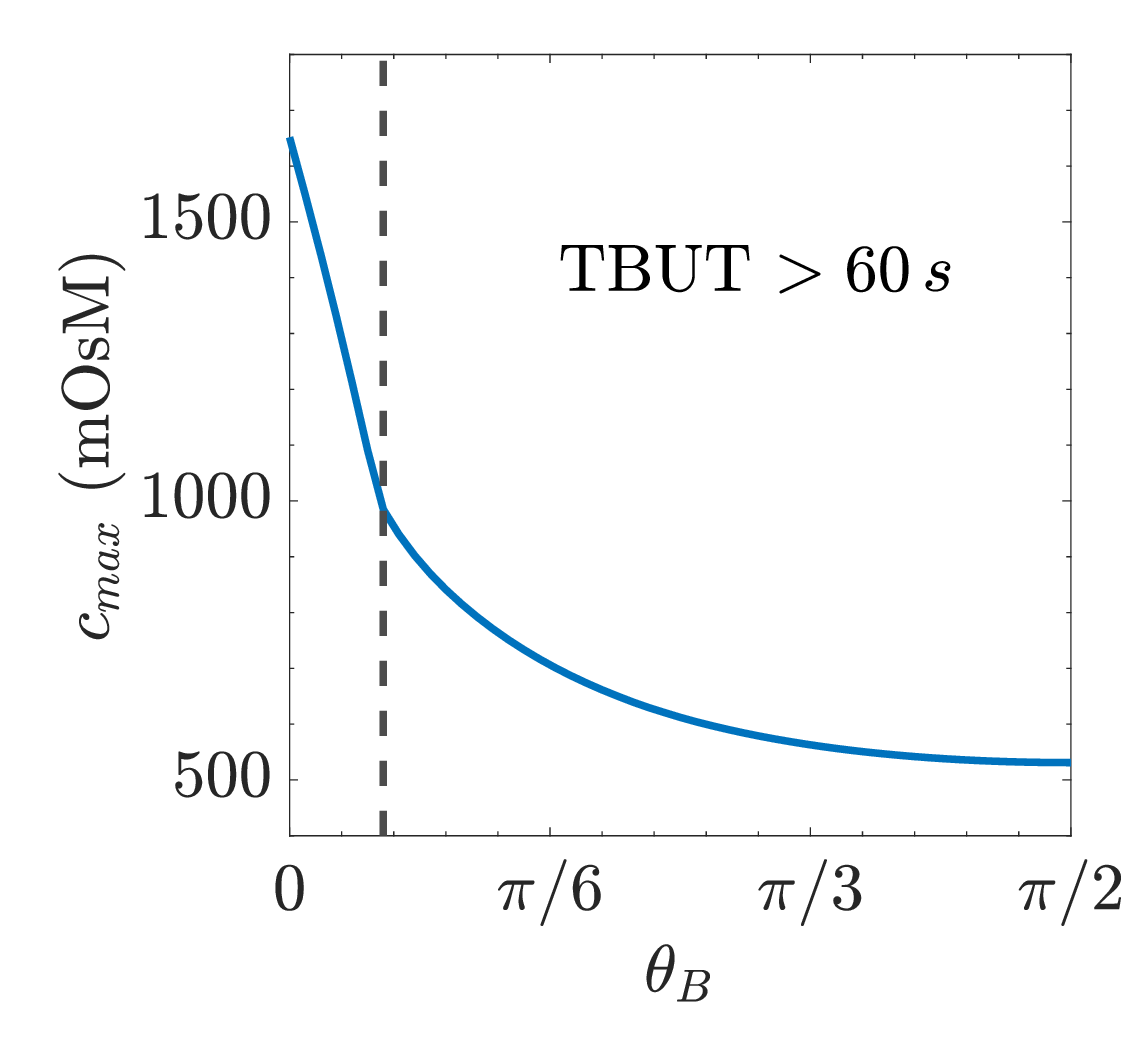}\\
\includegraphics[width=0.49\textwidth]{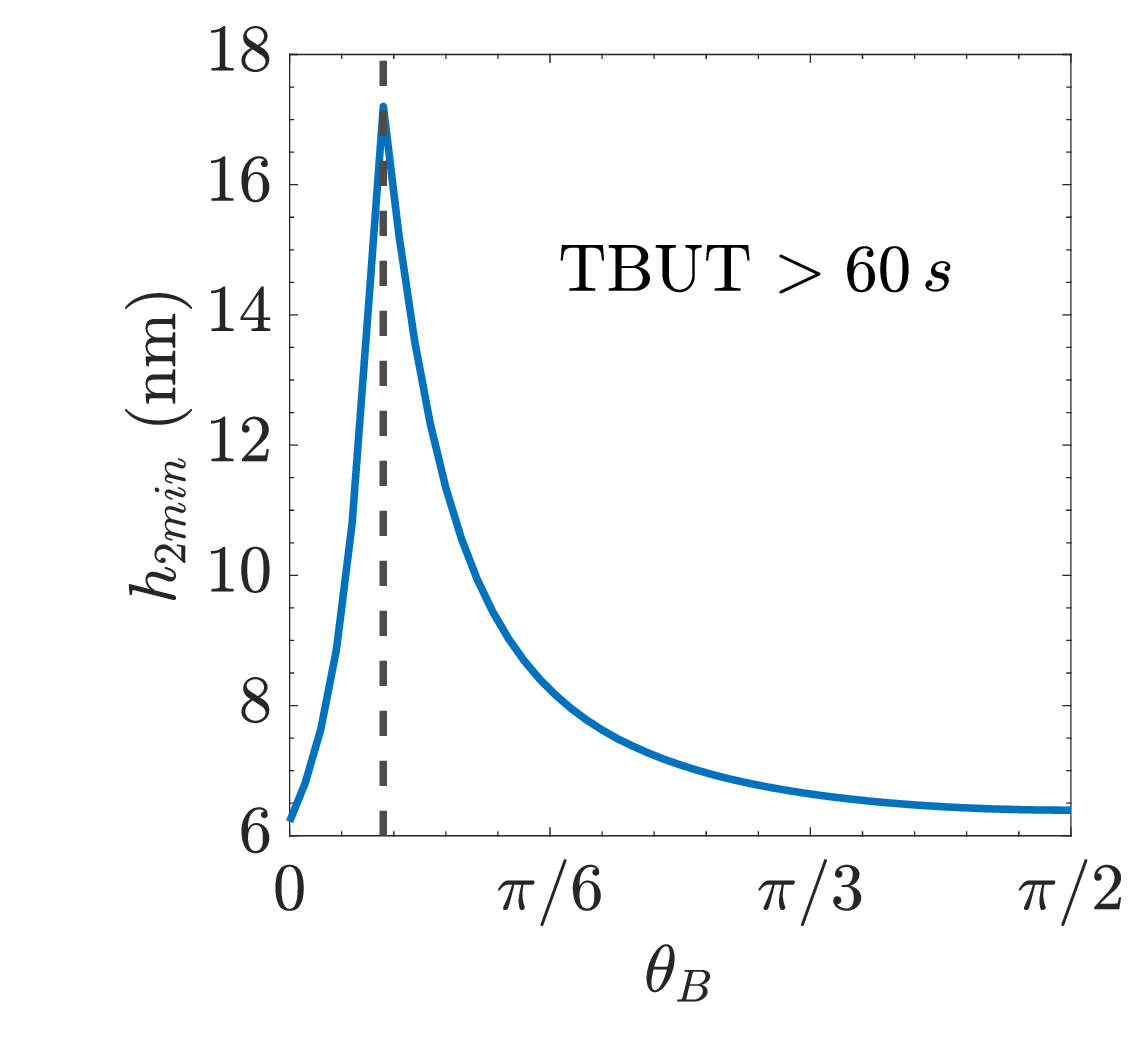}
\caption{Extreme values of AL thickness, osmolarity, and LL thickness at the end of each simulation as $\theta_B$ is varied from $0$ to $\pi/2$. To the left of the vertical dashed line at $\theta_B\approx \pi/17$, TBUT occurs before 60 s; to the right of this line, the simulations end at 60 s.  }\label{fig:extremavthetaB}
\end{figure}

While the overall TF dynamics are similar for our computations, some significant differences arise from varying the director angle (orientation) of the liquid crystal molecules.  We summarize the differences by plotting parametric changes to the minimum values of the AL and LL thicknesses, and the maximum value of osmolarity as $\theta_B$ varies from 0 to $\pi/2$. 
The values are shown at the final time of the simulation: at the TBUT if this occurs before 60 s, or at 60 s if TBU doesn't occur by that time.   Figure \ref{fig:extremavthetaB} shows TBUT results. For values of $\theta_B<\pi/17$, approximately, TBUT occurs before 60 s. In these cases, the minimum AL thickness reaches the threshold value of $0.5\mu$m. 
As $\theta_B$ increases to $\pi/17$, the minimum LL thickness increases and the LL thickens at its minimum; this represents a healing flow due to capillarity in the LL for this range of $\theta_B$. Peak osmolarity decreases, as the time to TBU lengthens, and the osmolarity has time to diffuse outward within the AL.

For $\theta_B>\pi/17$, TBU occurs after the final simulation time of 60 s. As the director angle increases, thinning in the AL slows, and the minimum AL thickness increases; by $\theta_B=\pi/2$, the minimum value at 60 s is $2.04\,\mu$m. From $0<\theta_B<\pi/6$, there is substantial movement in the LL. When $\theta_B$ is close to $\pi/17$, minimum thickness increases to over 17 nm. However, as $\theta_B$ increases from approximately $\pi/6$, the minimum LL thickness remains close to the initial minimum value of $5$ nm, increasing up to just 2 nm over the initial value. Peak osmolarity values continue to decrease as $\theta_B$ increases to $\pi/2$, as there is both more time for diffusion, and an increased aqueous volume which accompanies a lower peak osmolarity.  

\begin{figure}[htbp]%
\centering
\includegraphics[width=0.49\textwidth]{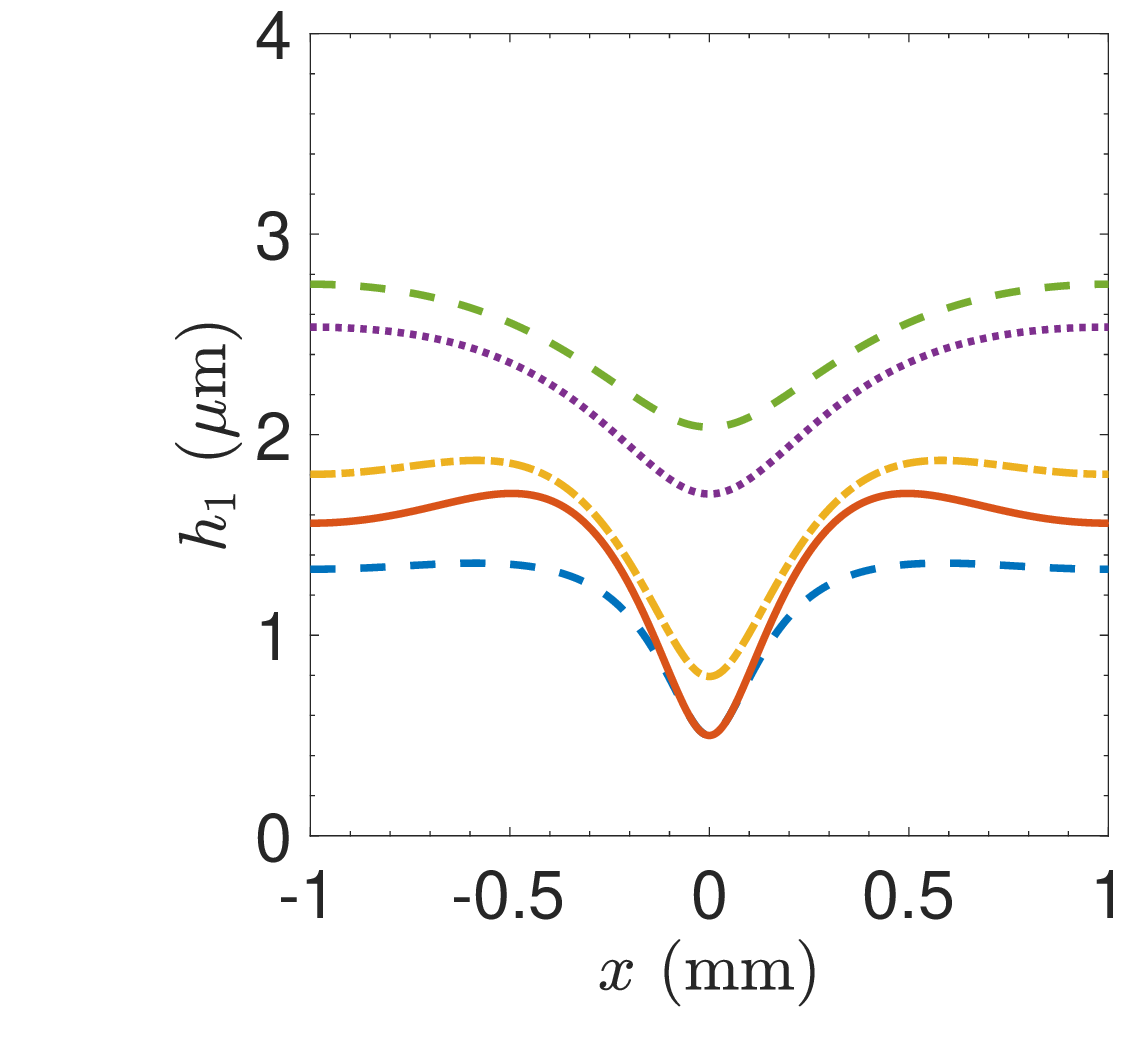}
\includegraphics[width=0.49\textwidth]{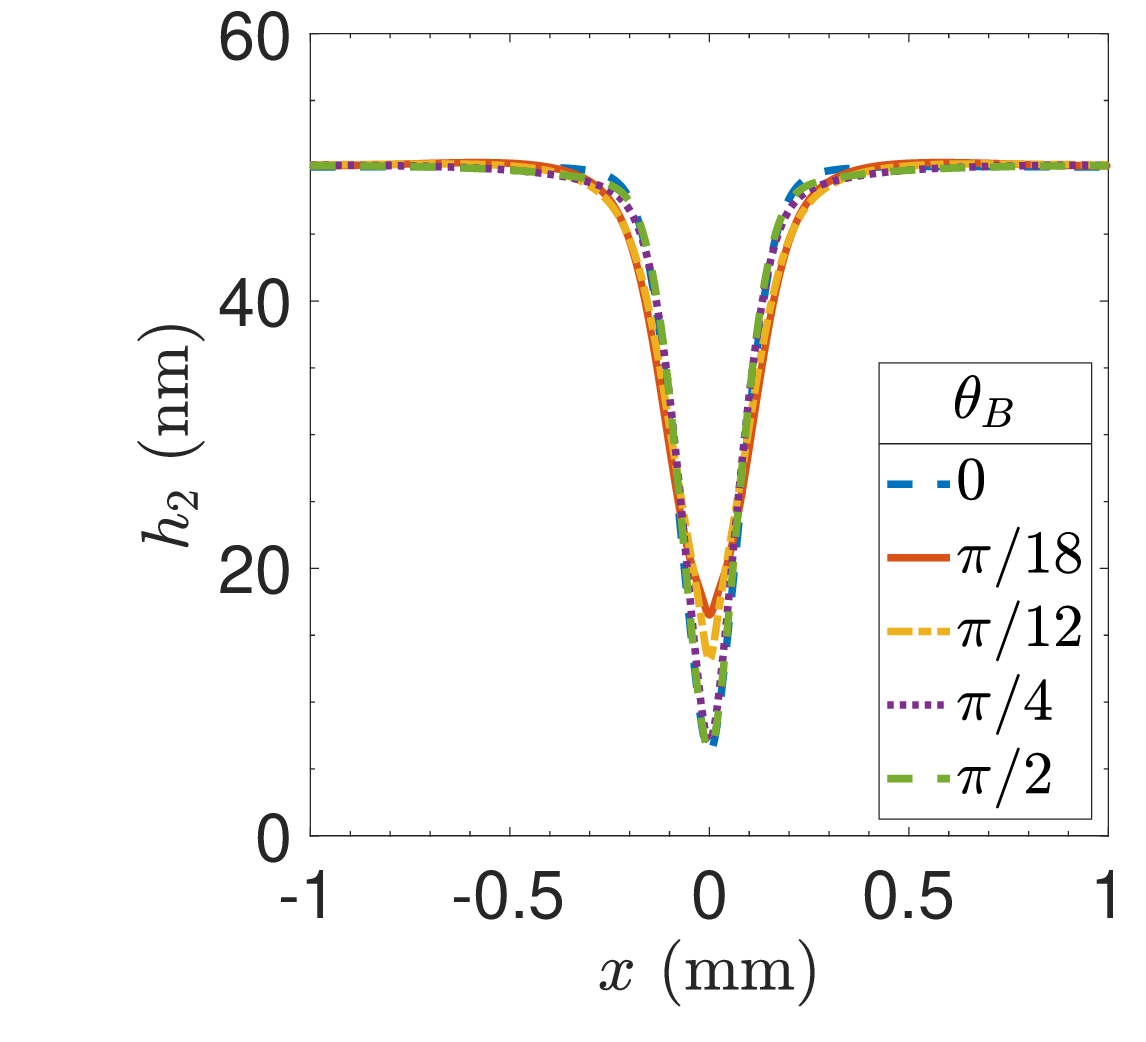}
\includegraphics[width=0.49\textwidth]{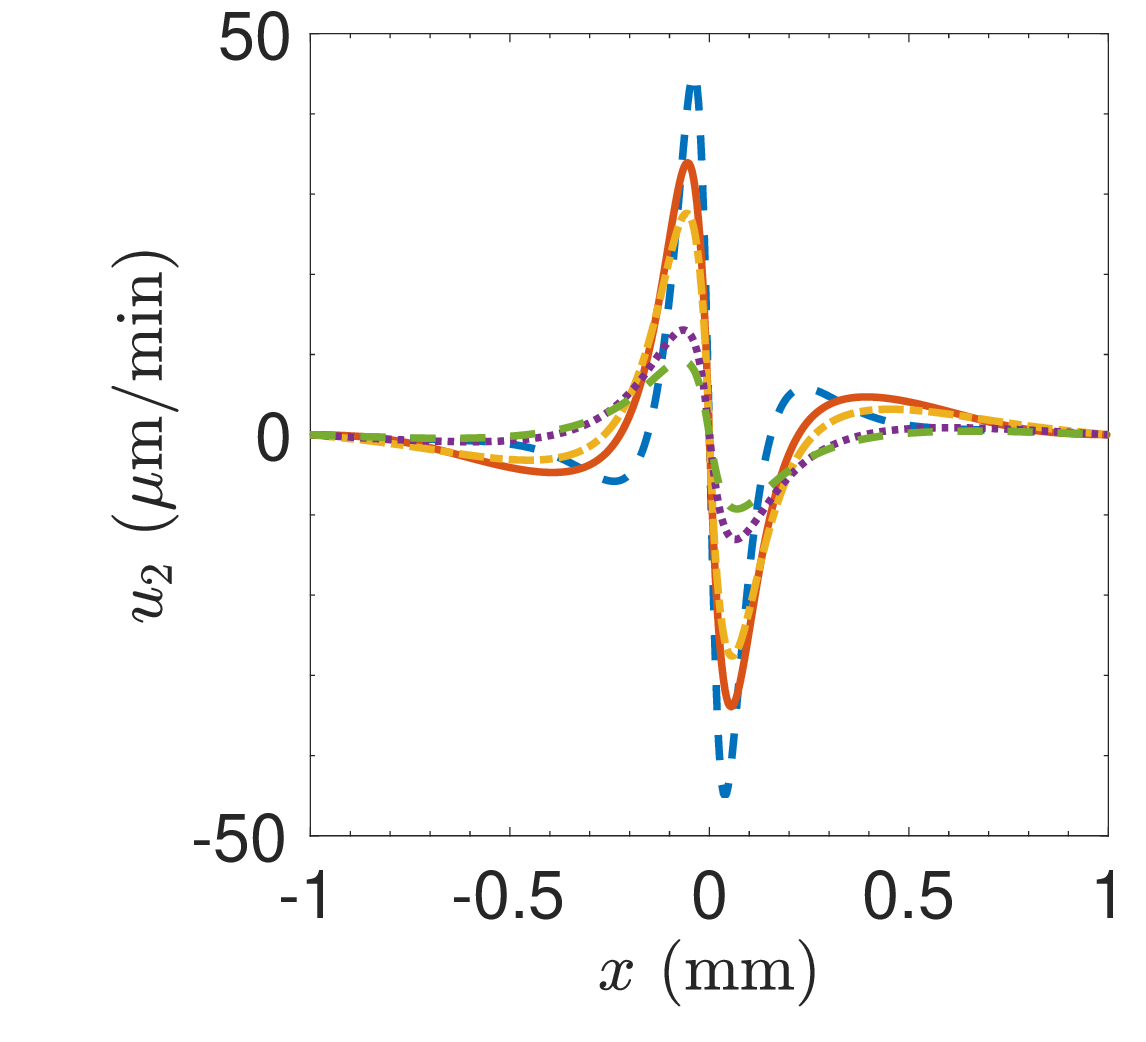}
\includegraphics[width=0.49\textwidth]{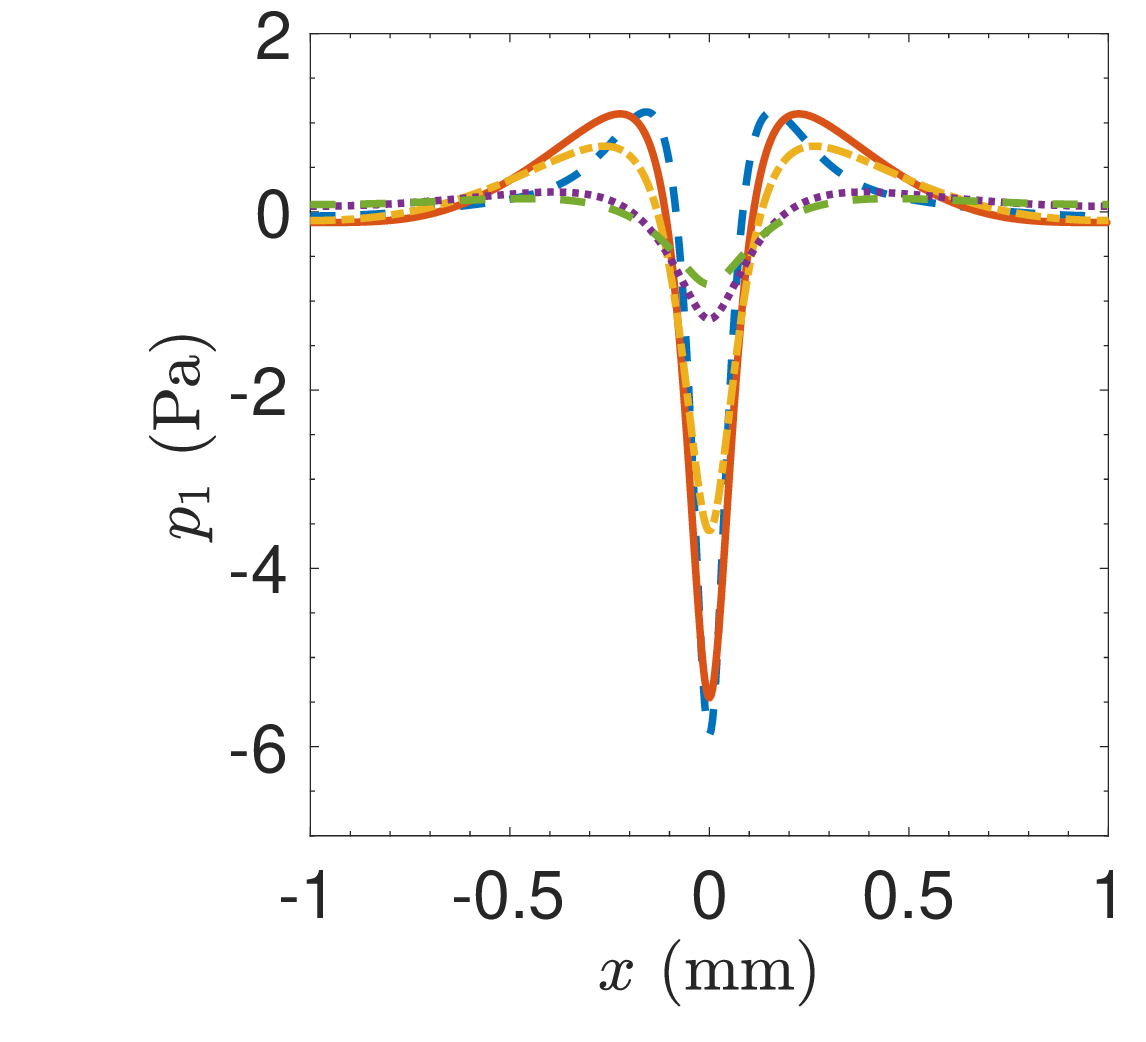}
\includegraphics[width=0.49\textwidth]{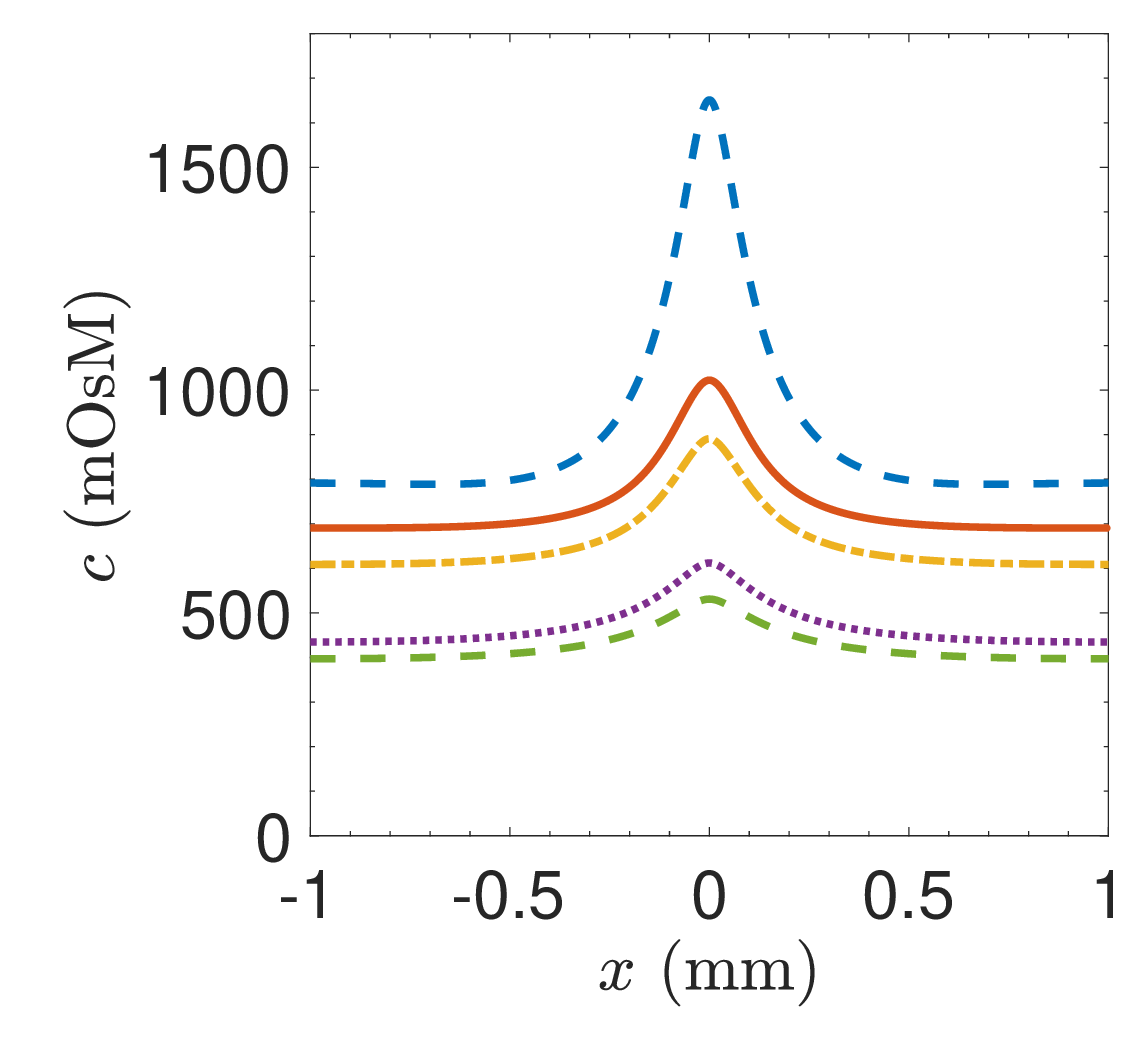}
\includegraphics[width=0.49\textwidth]{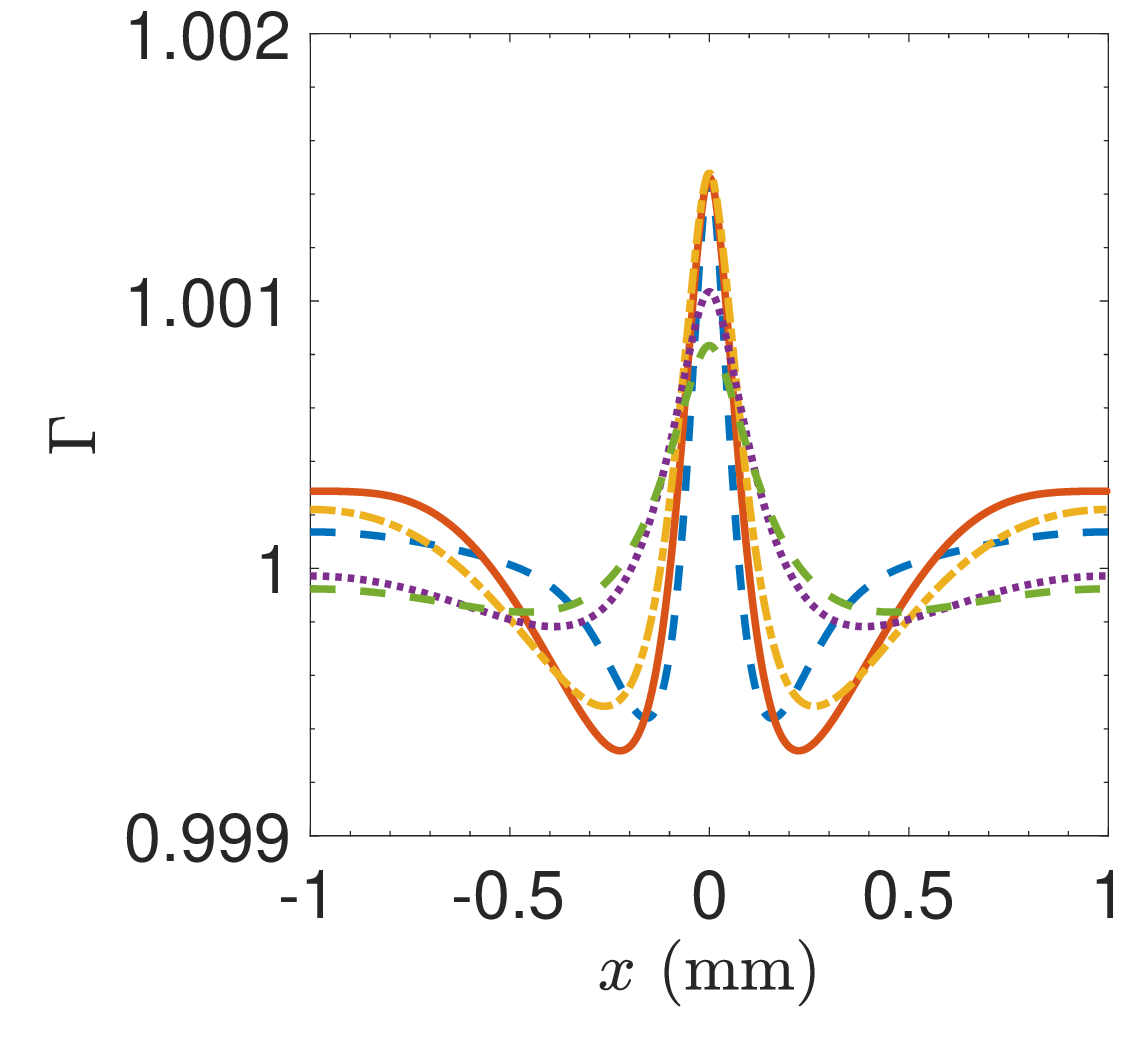}
\caption{Solutions at the final time; 15.9 s for $\theta_B=0$, 55.5 s for $\theta_B=\pi/18$, and 60 s for $\theta_B=\pi/12,\,\pi/4,$ and $\pi/2$.}\label{fig:finaltime}
\end{figure}

In summary, Figure \ref{fig:extremavthetaB} illustrates how orientation-dependence in the evaporation is coupled with flow. The lipid layer becomes much more mobile around $\theta_B=\pi/17$, indicating a mechanical response to the orientation of the liquid crystal molecules. The solutions at the final time for several values of $\theta_B$, shown in Figure \ref{fig:finaltime}, help to explain this effect in what follows.

\subsection{Solutions at the final time}

When $\theta_B=0$, the liquid crystal molecules are lying flat (parallel to a flat AL/LL interface). This orientation results in TBU at before 60 s. As in the base case ($\theta_B=\pi/2$, see Figure \ref{fig:basecase}), 
the aqueous layer develops a local minimum in the center of the domain, co-located with the LL minimum thickness. However, this valley is now more pronounced, and the entire layer thins relatively quickly.   The center reaches the threshold thickness of $0.5\,\mu$m by 15.9 s. The most pronounced AL local minimum occurs when $\theta_B=\pi/12$, although even by 60 s the TF has not approached TBU, with the minimum value just under 0.8 $\mu$m. For all the variables, the majority of variation occurs for small values of $\theta_B$; in particular, there is little difference in measurements between $\theta_B=\pi/4$ and $\theta_B=\pi/2$. 
The final thicknesses of the AL are summarized in Table \ref{table:max_osm}.

For many values of the director angle, except for those near $\theta_B=\pi/17$, the LL remains virtually tangentially immobile. Thus, the solution at the final time very closely resembles the initial condition. In contrast, for $\theta_B=\pi/12$, where the LL is the most mobile, the LL minimum increases as capillary-driven healing flow tries to fill it. 

For all values of $\theta_B$ studied here, thinning in the aqueous layer results in increased osmolarity, with peak osmolarities corresponding to the thinnest point in the aqueous layer. Initially, osmolarity is uniform in the aqueous layer at 300 mOsM. When $\theta_B=0$, TBU occurs at 15.9 s, the peak osmolarity is over five times the initial value at 1651.5 mOsM, and over two and a half times the initial value near the ends at 788.6 mOsM. For the other values of $\theta_B$, the peak in osmolarity is much smaller; see Table \ref{table:max_osm} for exact values.  

In all cases, surfactant forms a small pile in the center of the domain; the effect of the director angle is small, but the greatest variation occurs when $\theta_B=\pi/12$. 

\subsection{Extrema through time}
\begin{figure}[htbp]%
\centering
\includegraphics[width=0.49\textwidth]{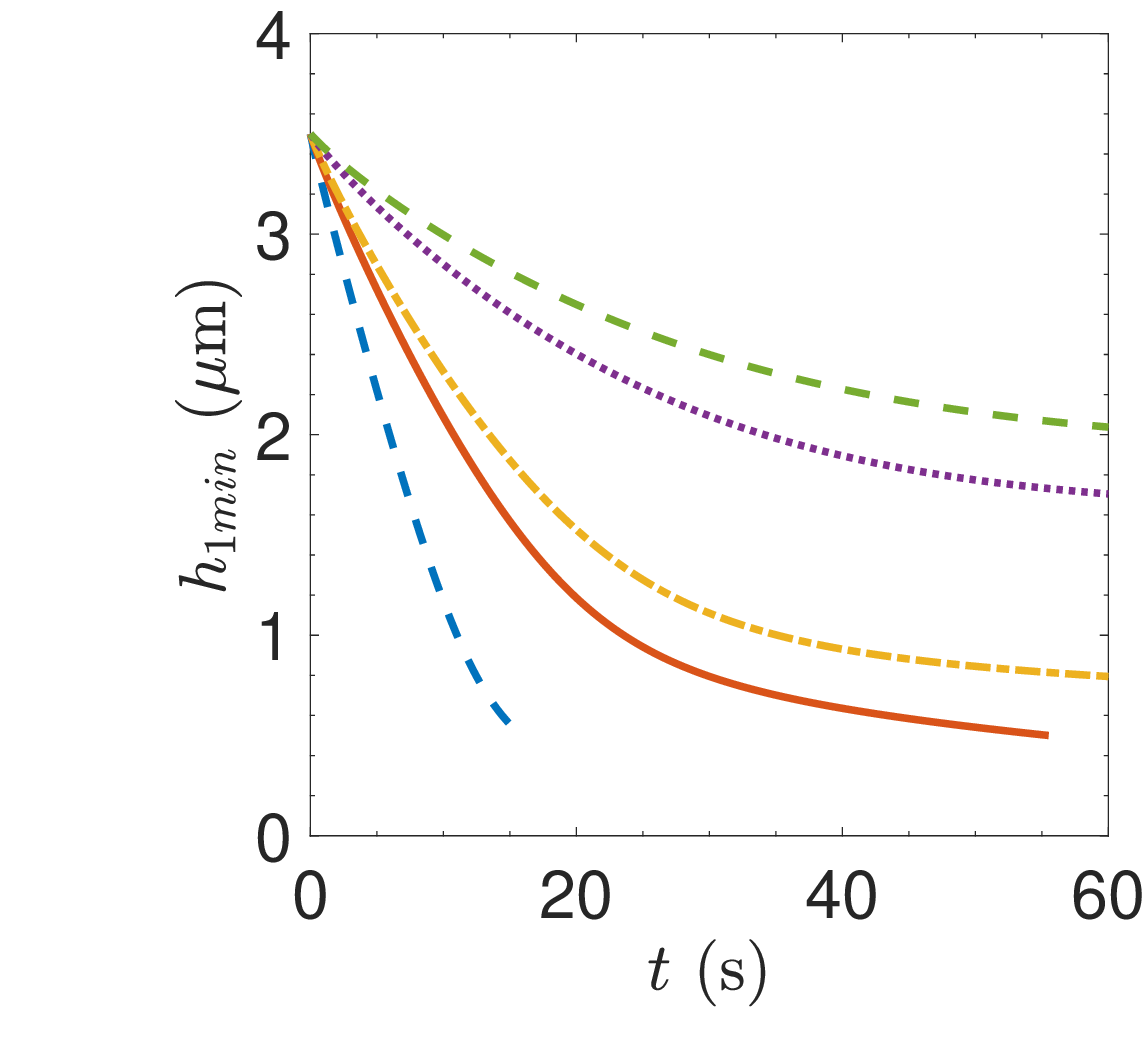}
\includegraphics[width=0.49\textwidth]{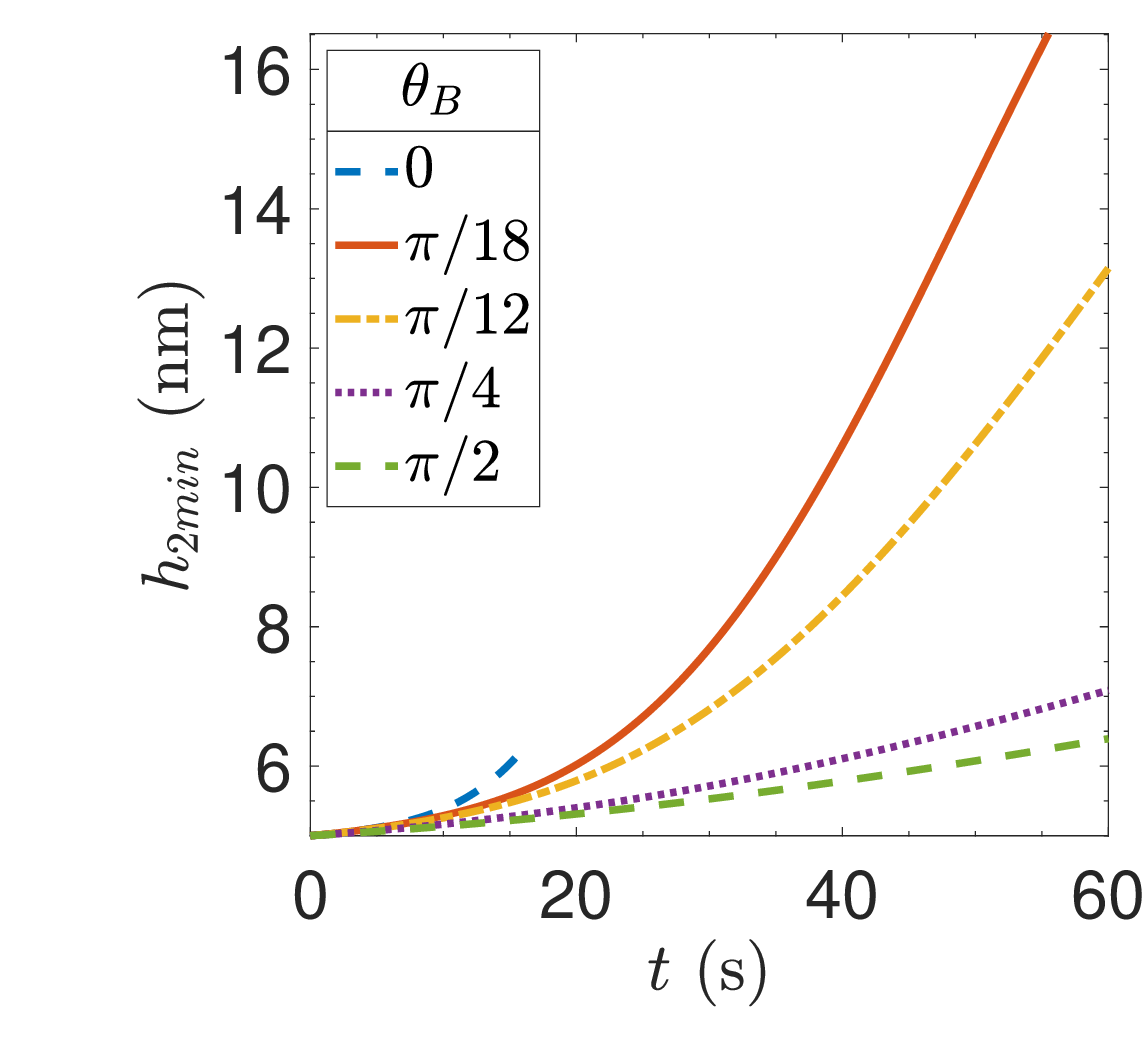}
\includegraphics[width=0.49\textwidth]{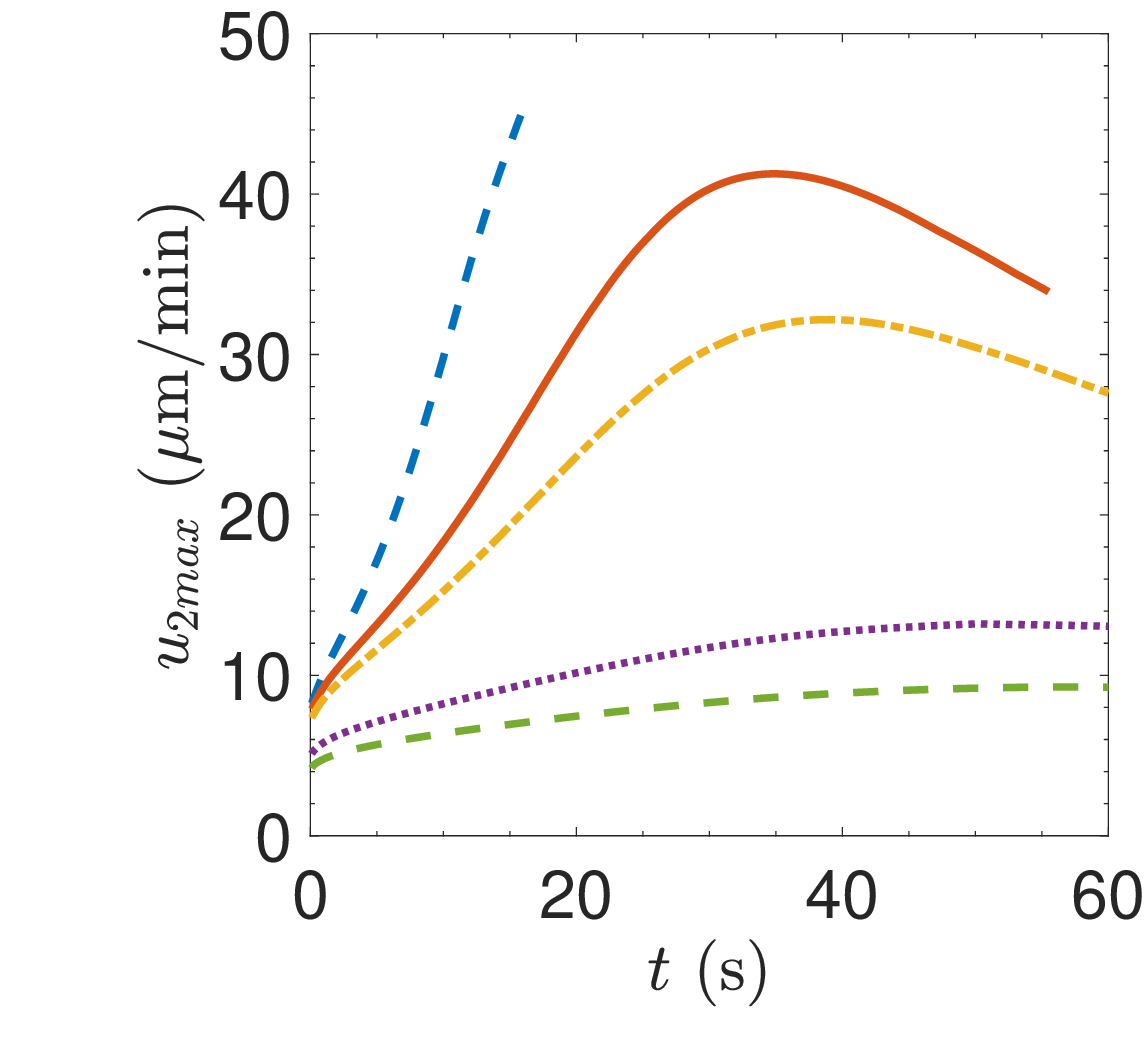}
\includegraphics[width=0.49\textwidth]{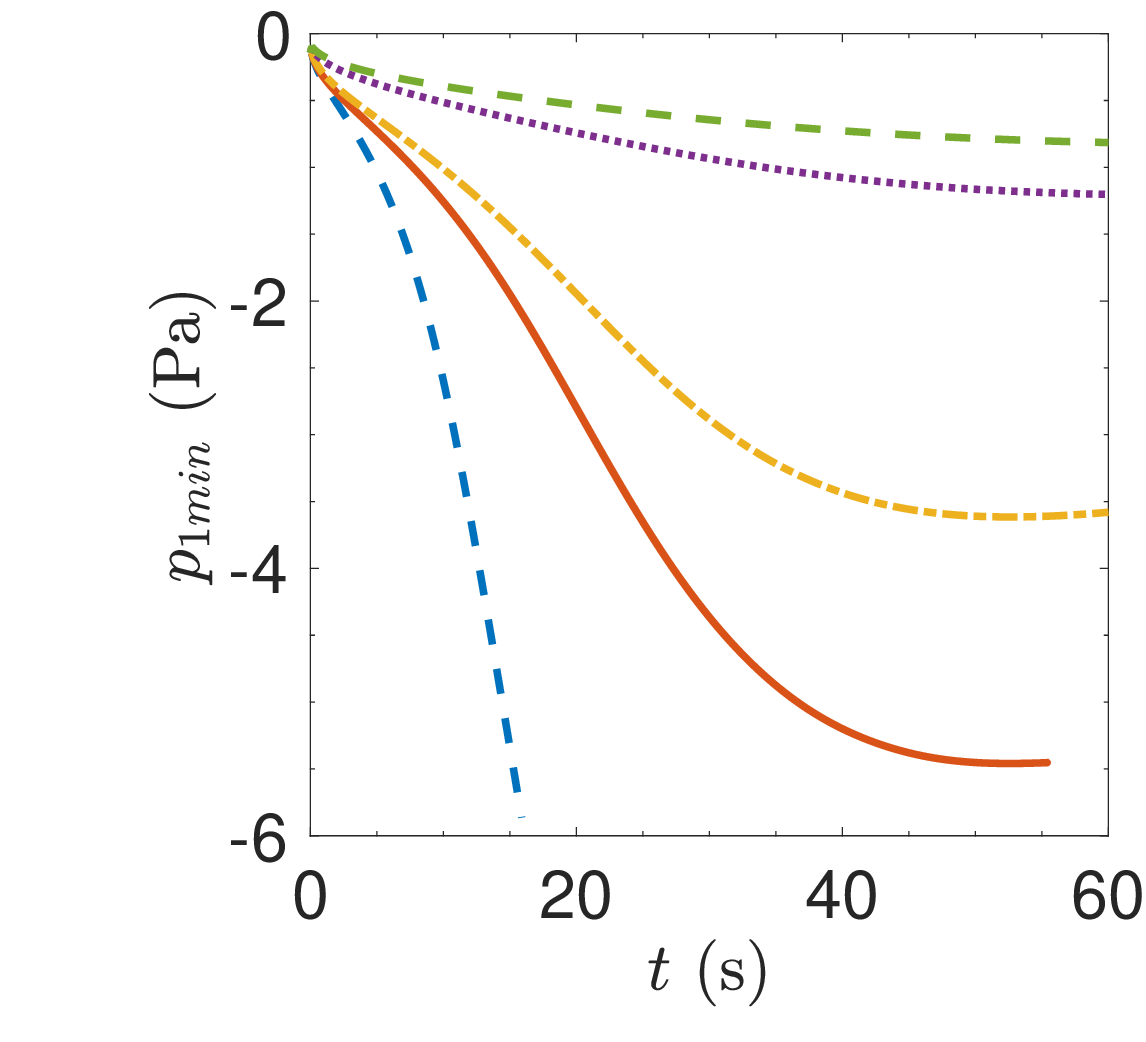}
\includegraphics[width=0.49\textwidth]{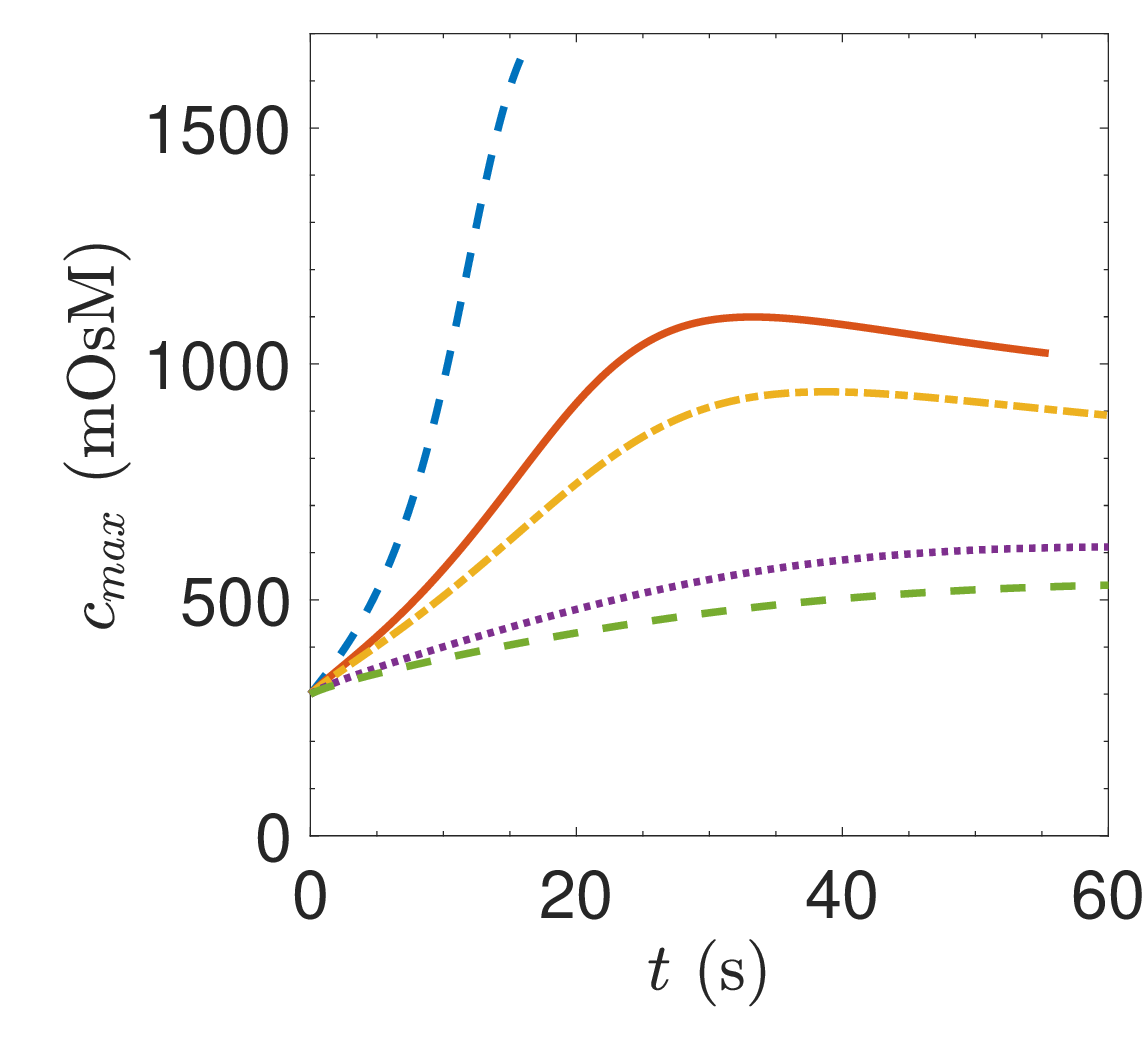}
\includegraphics[width=0.49\textwidth]{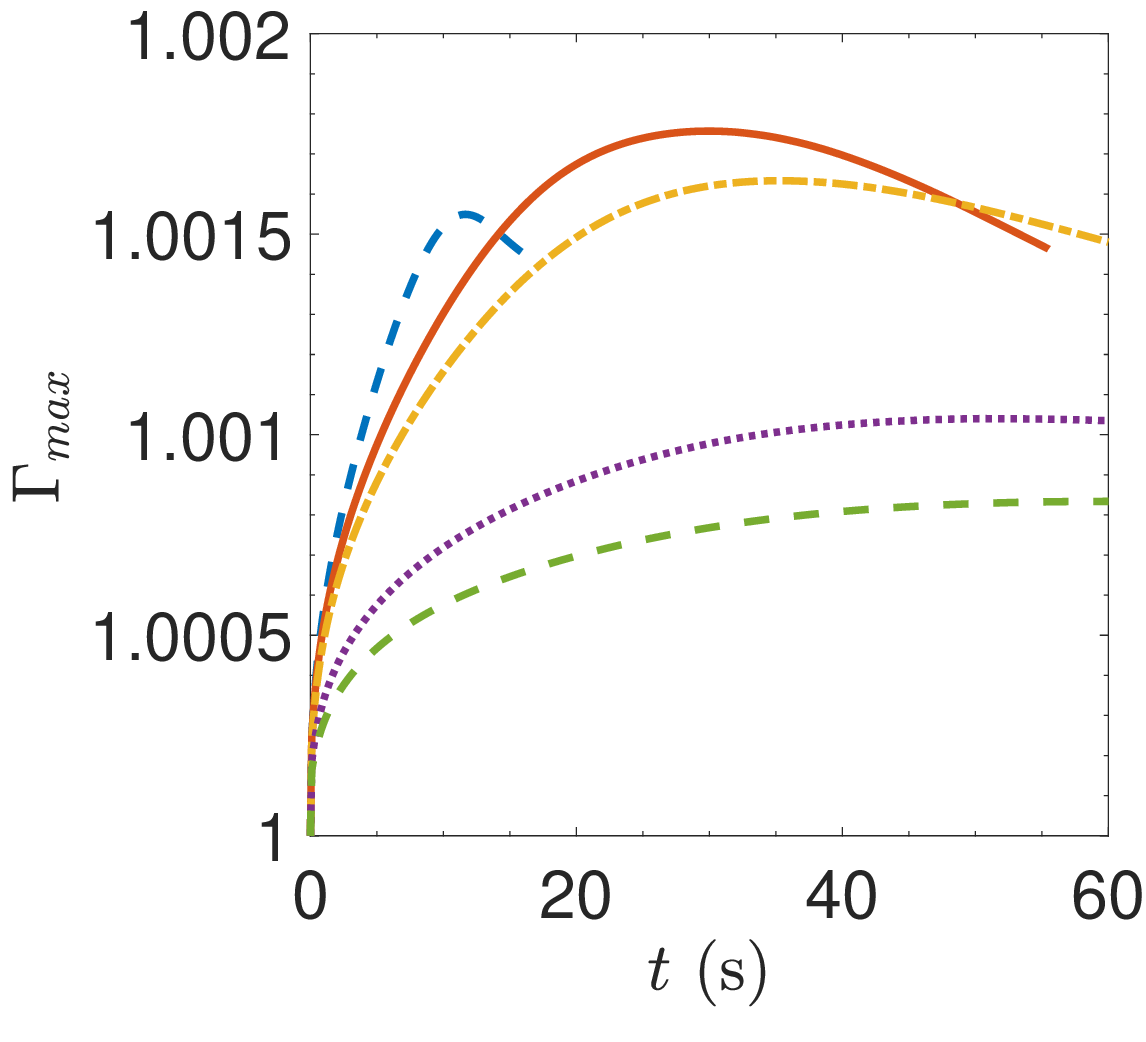}
\caption{Extrema through time for varying values of $\theta_B$. When $\theta_B=0$, TBU occurs at 15.9 s, and when $\theta_B=\pi/18$, TBU occurs at 55.5 s; for all other values, the simulation runs until the final time of 60 s. }\label{fig:extremathrutime}
\end{figure}

Having looked at the solutions at the final time, we now plot the time dependence of the extrema for the dependent variables; results appear in Figure \ref{fig:extremathrutime}. When $\theta_B=0$, the change in extrema is monotonic all variables except for  $\Gamma$, which peaks around 10 s and then decreases until TBUT is reached. When $\theta_B=\pi/4$ and $\pi/2$, the change in extrema is gradual. However, when $\theta_B=\pi/12$ and $\pi/18$, the dynamics during the first half of the simulation is generally different from the dynamics in the last half. On one hand, when $\theta_B=\pi/12$ the minimum AL thickness rapidly decreases from the initial value of $3.5\,\mu$m to $1.1\,\mu$m in the first 30 s but then slows, reaching a minimum of $0.79\,\mu$m at 60s. On the other hand, the minimum in the LL scarcely moves in the first 30 s, increasing by only $1.8$ nm, but then quickly increases reaching a minimum value of 13.1 nm at the final time. Peak osmolarity increases until about 35 s and then gradually decreases until the final time.  The decrease in peak osmolarity indicates that osmolarity is diffusing away from its maximum \cite{peng2014evaporation,BraunDrisTBU17}. Table \ref{table:max_osm} shows peak osmolarity values at 15, 30, and 60 s (except for $\theta_B=0,\,\pi/18$ which reach TBU before 60 s). 

\begin{table}[htbp]
\caption{Comparison of peak osmolarity levels and minimum AL thickness} \label{table:max_osm}%
\begin{tabular}{@{}llllcc@{}}
\toprule
 & \multicolumn{3}{c}{Max osmolarity (mOsM)}& Min AL thickness  & Final time\\
  $\theta_B$ value       &  15 s&  30 s& 60 s& ($\mu$m) & (s) \\
\midrule
         $0$&     1585.6&      -&    - &  0.5& 15.9\\
         $\pi/18$& 739.6& 1092.7&    - &  0.5& 55.5\\
         $\pi/12$& 626.4&  908.4&   891& 0.79& 60\\
         $\pi/4$&  441.7&  543.1& 612.1& 1.70& 60\\
         $\pi/2$&  404.3&  472.5&   531& 2.04& 60\\
\botrule
\end{tabular}
\end{table}

\section{Discussion and conclusion}
\label{sec:Discussion}
We expand upon previous two layer models of TBU in the tear film by representing the LL as a nematic liquid crystal, rather than as a fixed layer \cite{peng2014evaporation} or a Newtonian fluid \cite{stapf2017duplex}. We modify an existing evaporation model to account for the orientation of molecules in the LL as well as the dynamic thickness of the LL itself \cite{stapf2017duplex}. This enables us to explore the role of a hypothesized LL structure in preventing or slowing evaporation of the AL of the tear film. 

For some time, it is thought that the LL provided a physical barrier for evaporation \cite{MishimaMaurice61}, and thus a thicker LL corresponded to higher evaporative resistance. However, we now know that the thickness alone does not explain its effect; rather, it has been suggested that the evaporative resistance of the lipid layer also comes from its composition or structure \cite{king2010application,fenner2015moretostable,king-smith2015moretostable}.   

Experiments have shown that long, saturated hydrocarbon chains provide a good resistance to evaporation, and the longer the chain, the better the resistance \cite{archer1955rate}. Examination of the components in human meibum shows two types of molecules which fit this specification; namely the fatty acids of cholesterol esters, and the fatty alcohols of wax esters \cite{King-SmithOS13,paananen2019waxesters}. These molecules may organize to form lamellae in the LL with the hydrocarbon chains aligned with each other, though other interpretations are possible \cite{King-SmithOS13,rosenfeld2013structural,bland2019oahfa,viitaja2021tfll-and-oahfa}. Novel mixtures can now produce evaporation barriers in vitro \cite{svitova2021barrier,xu2023barrier}. Much work remains to be done on the link between structure and function in the LL.

By modeling the LL as a nematic liquid crystal in which the molecules have orientational but not positional order, we are able to incorporate some structure of the layer, which we believe may be key to understanding its function. Our hypothesized model shows that the orientation of the molecules in the lipid layer may have a noticeable effect on both the evaporation of the aqueous layer and the flow in the lipid layer. 

Nematic liquid crystals tend to align along the director angle $\theta_B$ which is constant in the limits used here. In our model, orientation has the largest effect when $\theta_B$ is small. When $\theta_B<\pi/17$, TBU occurs before the final simulation time of 60 s; for example, TBU occurs at 15.9 s for $\theta_B=0$, and 55.5 s when $\theta_B=\pi/18$. In these cases of TBU, maximum osmolarity at the time of breakup is over 1000 mOsM, which is well above the level associated with pain \cite{liu2009link}.

As $\theta_B$ is increased, the molecules in the lipid layer become more upright in orientation. This slows evaporation in the aqueous layer, by design, but also reduces movement in the lipid layer; the lipid layer remains tangentially immobile through the simulation time, changing very little from the initial condition. In this parameter regime, our model becomes similar to that hypothesized by \citet{peng2014evaporation} because of the LL immobility.  The director angle is also reminiscent of structures seen in vitro with analog molecules used to model the LL \cite{paananen2019waxesters,paananen2020CEs}.

The surfactant is also most mobile for small values of $\theta_B$. A small pile of surfactant builds up in the center of the domain, relaxing slightly before TBU. For larger values of $\theta_B$, TBU is not reached within the simulation time, and so the surfactant pile continues to build slowly until the end
time. Higher concentrations of surfactant act to pull fluid away, although we do not see this action because the healing flow from capillarity is larger for the parameters studied here. However, thinning can also be driven by surfactant; for example, a blob of surfactant in an otherwise initially-uniform LL could also cause TBU \cite{stapf2017duplex}. 

TBU and its relationship to DED have been analyzed in different approaches, namely by relying on neural networks \cite{vyasComprehensiveSurvey2020}. Some efforts aimed at quantifying TBUT measurement and related quantities  \cite{yedidyaAutoTBUSequenceDetect2008,ramosAnalysisParameters2014,suTearFilm2018}, while others aimed to use TBUT to diagnose DED \cite{remeseiroCASDESComputerAided2016,vyasDEDwithCNN2024}.  These approaches may well end up being useful in the clinic by automating current clinical practice but they do not aim to identify mechanisms underlying the TBU dynamics.  The approaches need not be separate and may yet combine to yield even more understanding of the relationship between TBU and DED; one such attempt found data automatically via neural network detection but then explored mechanisms driving TBU for a large number of instances in healthy subjects \cite{driscoll2023fittingMAIO}.  Such a combined method remains to be applied to DED subjects.

The goal of this paper was to derive a two layer model of the tear film which incorporates the LL as liquid crystal and includes molecular orientation in the evaporation function. By doing this we are able to explore possible ways in which the structure and composition of the lipid layer can affect tear film thinning and breakup. We have chosen to have the evaporation resistance depend on the orientation (director field), but we find that not only does this affect thinning in the aqueous layer which we expected, it also affects flow in the lipid layer, with a small angle of orientation causing the lipid layer to become more mobile. 

There are many parameters in this model. Future work will include a thorough parameter study. A two-layer model with the floating nematic LL approximated in the moderate elasticity limit would be a good future direction \cite{taranchuk2023extensional}. Including a moving end to approximate blinking in the eye would also be worthwhile. 

\bibliography{refs}

\end{document}